\documentclass[12pt,10pt,aps,pra,twocolumn]{revtex4-1}
\usepackage[latin9]{inputenc}
\usepackage{xcolor}
\usepackage{pdfcolmk}
\usepackage{amsmath}
\usepackage{graphicx}
\PassOptionsToPackage{normalem}{ulem}
\usepackage{ulem}

\makeatletter

\providecolor{lyxadded}{rgb}{0,0,1}
\providecolor{lyxdeleted}{rgb}{1,0,0}

\@ifundefined{textcolor}{}
{%
\definecolor{BLACK}{gray}{0}
\definecolor{WHITE}{gray}{1}
\definecolor{RED}{rgb}{1,0,0}
\definecolor{GREEN}{rgb}{0,1,0}
\definecolor{BLUE}{rgb}{0,0,1}
\definecolor{CYAN}{cmyk}{1,0,0,0}
\definecolor{MAGENTA}{cmyk}{0,1,0,0}
\definecolor{YELLOW}{cmyk}{0,0,1,0}
}

\makeatother

\begin{document}

\title{The universe on a table top: engineering quantum decay of a relativistic
scalar field from a metastable vacuum \\
}

\author{O. Fialko$^{1}$, B. Opanchuk$^{2}$, A. I. Sidorov$^{2}$, P. D.
Drummond$^{2}$, J. Brand$^{3}$ }

\affiliation{$^{1}$Dodd-Walls Centre for Photonic and Quantum Technologies, Institute
of Natural and Mathematical Sciences and Centre for Theoretical Chemistry
and Physics, Massey University, Auckland, New Zealand}

\affiliation{$^{2}$Centre for Quantum and Optical Science, Swinburne University
of Technology, Melbourne 3122, Australia}

\affiliation{$^{3}$Dodd-Walls Centre for Photonic and Quantum Technologies, New
Zealand Institute for Advanced Study and Centre for Theoretical Chemistry
and Physics, Massey University, Auckland, New Zealand}

\begin{abstract}
The quantum decay of a relativistic scalar field from a metastable
state (``false vacuum decay'') is a fundamental idea in quantum
field theory and cosmology. This occurs via local formation of bubbles
of true vacuum with their subsequent rapid expansion. It can be considered
as a relativistic analog of a first-order phase transition in condensed
matter. Here we expand upon our recent proposal {[}EPL 110, 56001
(2015){]} for an experimental test of false vacuum decay using an
ultra-cold spinor Bose gas. A false vacuum for the relative phase
of two spin components, serving as the unstable scalar field, is generated
by means of a modulated linear coupling of the spin components. We
analyze the system theoretically using the functional integral approach
and show that various microscopic degrees of freedom in the system,
albeit leading to dissipation in the relative phase sector, will not
hamper the observation of the false vacuum decay in the laboratory.
This is well supported by numerical simulations demonstrating the
spontaneous formation of true vacuum bubbles on millisecond time-scales
in two-component $^{7}$Li or $^{41}$K bosonic condensates in one-dimensional
traps of $\sim100\,\mu \mathrm{m}$ size.
\end{abstract}
\maketitle

\section{Introduction}

Bubble nucleation is a ubiquitous phenomenon in condensed matter physics~\cite{Schmelzer2005}.
The spontaneous creation of vapor bubbles due to thermal fluctuations
in superheated water and their collapse was studied more than 80 years
ago by Rayleigh in an attempt to explain sound emitted by a boiling
kettle~\cite{Rayleigh17}. Lifshitz and Kagan pioneered a quantum-mechanical
treatment of the first-order phase transition at zero temperature
through quantum nucleation of bubbles of a new phase~\cite{Lifshitz72}.
The quantum nucleation of bubbles was studied experimentally in $^{3}$He-$^{4}$He
mixtures~\cite{Satoh92}.

In a pioneering and inspirational theoretical study, Coleman subsequently
treated the quantum decay of a relativistic scalar field from a metastable
state, with formation of a true vacuum~\cite{Coleman1977}. Applied
to the universal inflaton quantum field, bubble nucleation is a model
for the cosmological ``big bang''~\cite{Vilenkin1983,Guth2007}.
Here bubbles nucleating from a false vacuum grow into universes, each
subsequently undergoing exponential growth of space~\cite{Coleman1980}.
Similar types of scenario are proposed for the development of particle
mass via the Higg's mechanism. This is fundamental to the current
standard model of particle physics. Understanding this process therefore
appears vital to the foundations of both cosmology and of particle
physics.

Although the concept is widely used in quantum field theory, Coleman's
theory was approximate, and confined to a thin-wall regime for the
scalar potential. Presently, no exact results are known for more general
potential landscapes. The decay of a relativistic false vacuum and
nucleation of a true vacuum has not been realized in any laboratory
experiment to test such theories. One obvious problem is the need
to have a system with a metastable potential for the internal potential
energy of the scalar field itself, a second is that the dynamics should
be driven by quantum fluctuations, not thermal noise, and a third
is the requirement of relativistic field dynamics. While qualitatively
analogous to bubble nucleation in condensed matter physics, this combination
of metastability, quantum fluctuations and relativistic dynamics makes
such models difficult to test quantitatively. An experiment in particle
physics, naturally desirable in principle, would require energies
far higher than those accessible using particle accelerators: and
it is almost unimaginable that the appropriate global initial conditions
would be available.

False vacuum decay, which initiates inflationary universe models,
is being tested against observations in astrophysical experiments
on the cosmic microwave background (CMB)~\cite{Feeney2011-inflation,Bousso2013}.
One difficulty is the need to disentangle gravitational effects from
quantum tunneling. This is not helped by the lack of a unified theory
of quantum gravity. From a theoretical point of view, quantum tunneling
from a false vacuum is a problem that has only been treated approximately~\cite{Coleman1977,Callan1977},
due to the exponential complexity of quantum field dynamics. This
motivates the search for an analog quantum system that is accessible
to experimental scrutiny, to test such models. The utility of such
experiments, which complement astrophysical investigations, is that
they would provide data that allow verification of widely used approximations
inherent in current theories.

In this paper we show how a relativistic false vacuum can be generated
with an ultra-cold atomic two-component spinor Bose-Einstein condensate
(BEC), extending upon our previous work~\cite{Fialko15}. The dynamics
of the dimensionless inflaton field $\phi$ in Coleman's model~\cite{Coleman1977}
is given by the equation
\begin{equation}
\partial_{t}^{2}\phi-c^{2}\nabla^{2}\phi=-\partial_{\phi}V(\phi),\label{eq:Hubble-1}
\end{equation}
where $c$ is the speed of light, and the effective potential $V(\phi)$
has a metastable local minimum separated from a true vacuum by a barrier.

The Coleman model is emulated in the BEC by the quantum field dynamics
occurring for the relative phase of two spin components that are linearly
coupled by a radio-frequency field. The speed of sound in the condensate
takes the role of the speed of light, and the spatial extent of the
``universe'' is less than a millimeter across. The true vacuum in
this system is the lowest energy state corresponding to the relative
phase being zero. The false vacuum is the metastable state with the
relative phase being $\pi$. The relevant initial state for the false
vacuum can be prepared by addressing a radio-frequency transition
between the spin components. Quantum decay from the false vacuum is
expected to seed the nucleation of bubbles, which are spatial regions
of true vacuum. These bubbles can be observed interferometrically~\cite{Egorov2011}
over millisecond time-scales. While a radio-frequency coupling between
the spin components with constant amplitude can create an unstable
vacuum~\cite{Opanchuk2013-early-universe}, an amplitude modulation
in time allows one to create a metastable vacuum~\cite{Fialko15}.
The principle behind the vacuum stabilization is identical to the
one of stabilizing the unstable point of a pendulum by rocking the
pivot point as suggested by Kapitza~\cite{Kapitza1951-JETP}. For
a related recent application of this idea see~\cite{Citro2015a}.

Our proposal requires a two-component BEC where repulsive intra-component
interactions dominate over inter-component interactions. We have identified
a Feshbach resonance of $^{41}$K for scattering between two hyper-fine
states with a zero crossing for the inter-component $s$-wave scattering.
We expect that other candidate systems may exist as well. An alternative
implementation of the model in 1 or 2 space dimensions could also
be achieved with a single-species BEC and a double-well potential
in the tight-binding regime, where the tunnel-coupling is modulated
in time by changing the trap parameters.

In addition to the quantum field dynamics of the relative phase of
the spin components, there is a coupling to phonon degrees of freedom
in our system, which serves to damp the dynamics~\cite{Caldeira1981}.
Our studies suggest that damping can be reduced by an appropriate
choice of the experimental parameters, showing the feasibility of
a table-top experiment.

Analog models of the early Universe with ultra-cold atoms have previously
been considered in the literature with a focus on different phenomena,
including the inflationary expansion of space-time~\cite{Fischer2004,Menicucci2010},
the formation of long-lived localized structures~\cite{Neuenhahn2012-phase-structures,Su2014a},
and the decay from an unstable vacuum~\cite{Opanchuk2013-early-universe}.
In this paper we present an analog model of false vacuum quantum decay.
This is relevant to the early quantum nucleation stage of bubbles
where gravitational effects are irrelevant even in cosmological models~\cite{Coleman1980}.
By contrast, the gravitationally dominated later stages of cosmological
evolution like bubble growth, slow-roll inflation, and re-heating,
can be simulated efficiently on computers due to their largely classical
nature~\cite{Liddle2000,Amin2012}. Our model is particularly interesting
in that it potentially allows an experimental test of quantum tunneling
in the regime of a relatively broad well, relevant to an inflationary
universe scenario, rather than the thin-wall potentials required for
the application of the Coleman instanton approximation.

Our Letter~\cite{Fialko15} has previously introduced the model,
focussing on vanishing inter-component interaction, and demonstrated
the stabilization, and metastable properties of the false vacuum by
analytic methods and numerical simulations. In this paper we describe
and analyze the model in more detail. First, we derive rigorously
the effective time-independent Hamiltonian of the system by applying
time-dependent perturbation theory to a driven two-component BEC (sections~\ref{subsec:Many-Body-Kapitza-Pendulum}
and~\ref{subsec:Effective-Hamiltonian}). Then we analyze the dynamics
of the model within a functional integral approach, supplanting a
less rigorous analysis performed in Ref.~\cite{Fialko15}, in section~\ref{sec:Functional-Integral-representati}.

In section~\ref{sec:Tunneling-rate} we use Coleman's analytical
approach to approximately calculate the tunnelling rate and arrive
at a new scaling law~\eqref{eq:B1D}, which restricts the possible
exponent functions. The scaling law obtained here is substantiated
by numerical simulations of the quantum field dynamics of the spinor
BEC in the truncated Wigner approximation (TWA)~\cite{Drummond1993a,Steel1998,Sinatra2002}
in section~\ref{subsec:Symmetric-BEC-experiment}, which also gives
quantitative results for the scaling exponents that can be experimentally
tested. This numerical approach relies on completely different approximations
to the analytic theory, and has been widely applied and quantitatively
tested in experiments in one and higher dimensions for both interacting
photons and ultra-cold atoms~\cite{Drummond2016}. It is known to
agree with exact simulations of simplified models of quantum tunneling
dynamics~\cite{Drummond1989} in the important near-threshold regime
where the Coleman approximation may not be applicable.

Section~\ref{subsec:Non-zero-inter-component-interac} generalizes
the model equations to non-zero inter-component interactions and energy
calculations in section~\ref{subsec:Energy-calculations} provide
insight into the damping of the decay dynamics by leaking of energy
into the phonon sector. The experimental procedure for realizing the
model with $^{41}$K atoms is detailed in section~\ref{sec:Experimental-proposal},
where also the situation with $^{7}$Li atoms is briefly considered.

\section{The model\label{sec:The-model}}

We consider a two-component BEC of atoms with mass $m$ and with a
time-dependent coupling $\nu+\delta\hbar\omega\cos(\omega t)$ between
two components. Atoms with the same spin interact via a point-like
potential with strength $g_{jj}$ (here $j$ is either $1$ or $2$),
while atoms with different spin components interact via a point-like
potential with strength $g_{12}$. The Hamiltonian of the system is

\begin{equation}
\begin{split}\hat{H}= & \int d{\bf r}\hat{\psi}_{j}^{\dagger}\left[-\frac{\hbar^{2}\nabla^{2}}{2m}-\mu\right]\hat{\psi}_{j}+\frac{g_{jk}}{2}\int d{\bf r}\hat{\psi}_{j}^{\dagger}\hat{\psi}_{k}^{\dagger}\hat{\psi}_{k}\hat{\psi}_{j}\\
- & [{\normalcolor \nu+\delta\hbar\omega\cos(\omega t)]}\int d{\bf r}\hat{\psi}_{j}^{\dagger}\hat{\psi}_{3-j},
\end{split}
\label{eq:Hamiltonian}
\end{equation}
where summation over spin indices $j=1,2$ and $k=1,2$ is implied.
The Bose fields satisfy the usual commutation relations $\left[\hat{\psi}_{j}({\bf r}),\hat{\psi}_{k}^{\dagger}({\bf r}')\right]=\delta_{jk}\delta({\bf r}-{\bf r'})$.
The chemical potential $\mu$ has no physical significance here, but
sets the energy scales.

This model can be formulated in up to three space dimensions.
The intra-component coupling $\nu$ is due to an imposed microwave
field that couples two hyperfine levels in an external magnetic field.
This is a standard effective low-energy Hamiltonian used to describe
many recent experiments in ultra-cold atomic physics below the Bose-condensation
temperature~\cite{PhysRevA.80.023603}.

The frequency $\omega$ represents an additional amplitude modulation
frequency of the microwave field, causing the coupling to vary sinusoidally
with time at a frequency much lower than the microwave carrier frequency.
The constant $\delta$ is a dimensionless quantity that gives the
depth of modulation in energy units of $\hbar\omega$.

\subsection{Many-Body Kapitza Pendulum\label{subsec:Many-Body-Kapitza-Pendulum}}

To gain insight into the physics of the modulated coupling, we consider
the behavior of the relative phase degree of freedom in the semi-classical,
mean-field, homogeneous limit. In the simplest case that $g_{12}=0$,
$g=g_{11}=g_{22}>0$, with $\mu=g\rho_{0}+\nu$, the lowest energy
manifold is for equal densities in the two spin components.

We now consider the classical equation of motion for the relative
phase $\phi_{a}=\phi_{1}-\phi_{2}$ prior to turning to the effective
Hamiltonian picture. Here we define $\psi_{j}=\sqrt{\rho_{j}}\exp(i\phi_{j})$,
so that if we assume that the density is equal and constant for the
two components, with $\rho_{j}=\rho_{0}$, then the relative phase
evolves according to

\begin{equation}
\partial_{t}^{2}\phi_{a}=-\frac{4g\rho_{0}}{\hbar^{2}}[\nu+\delta\hbar\omega\cos(\omega t)]\sin(\phi_{a}),\label{eq:Pendulum}
\end{equation}
which describes the movement of a periodically-driven pendulum.

We consider fast modulations of the coupling with frequency $\omega$
higher than any internal characteristic frequency in the system. According
to Kapitza~\cite{Kapitza1951-JETP} the ``angle'' $\phi_{a}$ may
be viewed now as a superposition $\phi_{a}=\phi_{0}+\Xi$ of a slow
component $\phi_{0}$ and a rapid oscillation $\Xi$.

Substituting this into the equation of motion and keeping the largest
terms we extract $\Xi=\delta\hbar\omega_{0}^{2}/(\omega\nu)\sin(\phi_{0})\cos(\omega t)$.
We then substitute $\phi_{a}=\phi_{0}+\Xi$ with the known $\Xi$
into the equation of motion to find an equation for $\phi_{0}$, keeping
terms up to first order in $\omega^{-1}$ and averaging over rapid
oscillations in time.

The resulting equation of motion for $\phi_{0}$ is
\begin{equation}
\partial_{t}^{2}\phi_{0}=-\partial_{\phi_{0}}V(\phi_{0})
\end{equation}
\begin{figure}[!t]
\includegraphics{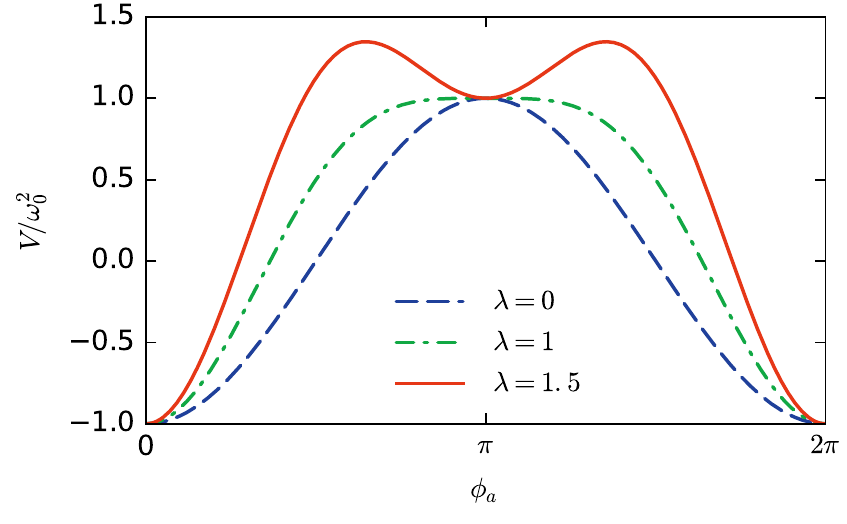}

\caption{\textcolor{black}{Effective field potential $V$ given in Eq.~\eqref{eq:Potential}
for different values of $\lambda$. It develops a local minimum at
$\phi_{a}=\pm\pi$ for $\lambda>1$. We take $\omega_{0}=1$ here
for purposes of illustration.\label{fig:Effective-field-potential}}}
\end{figure}
Here, the potential $V(\phi_{0})$ is plotted in Fig~\ref{fig:Effective-field-potential},
and is given analytically by
\begin{equation}
V(\phi_{a})=-\omega_{0}^{2}\left[\cos(\phi_{a})-\frac{\lambda^{2}}{2}\sin^{2}(\phi_{a})\right],\label{eq:Potential}
\end{equation}
where we have defined a characteristic frequency $\omega_{0}$ due
to the coupling as:

\begin{equation}
\omega_{0}=2\sqrt{\nu g\rho_{0}}/\hbar\label{eq:Omega0}
\end{equation}
 and a dimensionless parameter $\lambda$ that parameterizes the depth
of Kapitza modulation

\begin{equation}
\lambda^{2}=2\rho_{0}g\delta^{2}/\nu.\label{eq:Lamnda}
\end{equation}

\begin{figure}[!t]
\includegraphics{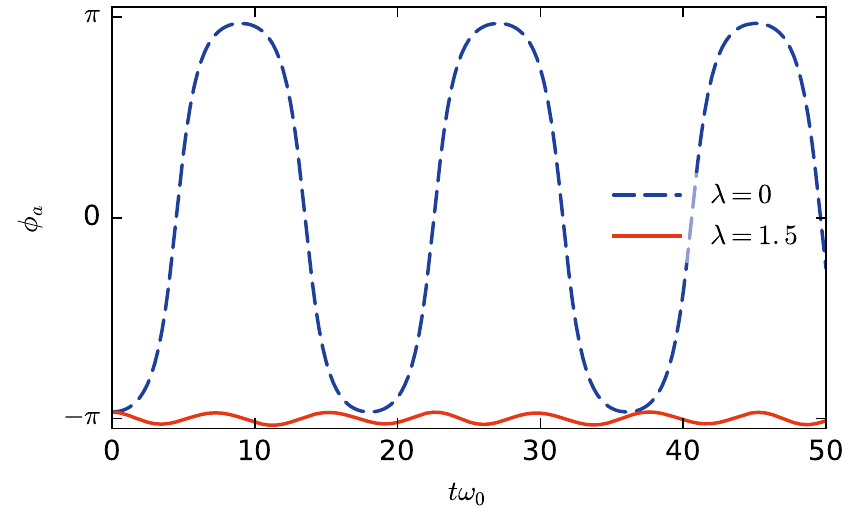}

\caption{\textcolor{black}{Solutions of Eq.~\eqref{eq:GP} for the relative
phase $\phi_{a}$, for different values of the parameter $\lambda^{2}=2\delta^{2}g\rho_{0}/\nu$.
}The initial state is chosen to be $\psi_{1}=\sqrt{\rho_{0}}$ and
$\psi_{2}=\sqrt{\rho_{0}}\exp(i\pi+i0.1)$\textcolor{black}{. For
purposes of illustration, we use dimensionless parameters with $\hbar=1$,
$g_{12}=0$, $g_{11}=g_{22}=g$, $\mu=g\rho_{0}$+$\nu$, $\rho_{0}=100$,
$\nu=0.1g\rho_{0}$, $\omega=50\omega_{0}$. For $\lambda=1.5$, in
the red line, the effective potential for the relative phase develops
a local minimum at $\phi_{a}=-\pi$, cf Fig.~\ref{fig:Effective-field-potential}.
As a result, the phase oscillates around the new local equilibrium
value, while at $\lambda=0$ it freely rolls down the hill.\label{fig:Kapitza}}}
\end{figure}

To demonstrate the resulting many-body Kapitza pendulum, we solve
the time-dependent coupled mean-field Gross-Pitaevskii equations obtained
from Eq.~\eqref{eq:Hamiltonian}

\begin{eqnarray}
i\hbar\partial_{t}\psi_{j} & = & \left[-\frac{\hbar^{2}}{2m}\nabla^{2}-\mu+g_{jj}|\psi_{j}|^{2}+g_{12}|\psi_{3-j}|^{2}\right]\psi_{j}\nonumber \\
 & - & [\nu+\delta\hbar\omega\cos(\omega t)]\psi_{3-j},\label{eq:GP}
\end{eqnarray}
where index $j=1,2$. Oscillations of the relative phase $\phi_{a}$
for two different values of the modulation parameter $\lambda$, for
a uniform field, are shown in Fig.~\ref{fig:Kapitza}.

\subsection{Effective Hamiltonian\label{subsec:Effective-Hamiltonian}}

We now consider how to translate this semi-classical result into a
quantum dynamical equation, for the general coupling case with arbitrary
$g_{ij}$. For such fast oscillations we derive an effective time-independent
Hamiltonian ruling the time average dynamics following the approach
developed in Ref.~\cite{goldman14}. For a single-harmonic modulation
of the form $\hat{H}=\hat{H}_{0}+\hat{H}_{1}\cos(\omega t)$, the
effective Hamiltonian reads

\begin{equation}
\hat{H}_{{\rm eff}}=\hat{H}_{0}+\frac{1}{4(\hbar\omega)^{2}}\left[\left[\hat{H}_{1},\hat{H}_{0}\right],\hat{H}_{1}\right]+{\cal O}(1/\omega^{3}).\label{eq:Correction0}
\end{equation}
Here, $\hat{H}_{0}$ is the same as $\hat{H}$ without driving. Using
$\hat{H}_{1}=-\delta\hbar\omega\int d{\bf r}\hat{\psi}_{j}^{\dagger}\hat{\psi}_{3-j}$
, the second term in Eq.~\eqref{eq:Correction0} reads

\begin{eqnarray}
 & - & \frac{(g_{jj}-g_{12})\delta^{2}}{4}\int d{\bf r}\left[\left(\hat{\psi}_{j}^{\dagger}\right)^{2}\left(\hat{\psi}_{3-j}\right)^{2}+\left(\hat{\psi}_{3-j}^{\dagger}\right)^{2}\left(\hat{\psi}_{j}\right)^{2}\right.\nonumber \\
 & + & \left.2\hat{\psi}_{j}^{\dagger}\hat{\psi}_{j}^{\dagger}\hat{\psi}_{j}\hat{\psi}_{j}-4\hat{\psi}_{j}^{\dagger}\hat{\psi}_{j}\hat{\psi}_{3-j}^{\dagger}\hat{\psi}_{3-j}\right].\label{eq:Correction}
\end{eqnarray}
The harmonic modulation thus leads to two-particle tunneling processes
encapsulated in the first line of Eq.~\eqref{eq:Correction} as well
as to the modification of the interaction strengths encapsulated in
the second line of Eq.~\eqref{eq:Correction}. To shed light on its
nature we ignore density fluctuations and represent the fields as
$\psi_{j}\approx\sqrt{\rho_{0}}\exp(i\phi_{j})$, where $\rho_{0}$
and $\phi_{j}$ are the density and the phase of a single component,
recalling that we are considering here the symmetric case where both
components have equal density, apart from quantum density fluctuations
that are neglected at this stage.

Substituting this into Eq.~\eqref{eq:Correction} and combining with
the single tunneling term $-\nu\int d{\bf r}\hat{\psi}_{j}^{\dagger}\hat{\psi}_{3-j}$
we obtain the mean-field potential energy $U$ felt by the relative
phase $\phi_{a}=\phi_{1}-\phi_{2}$

\begin{equation}
U(\phi_{a})\approx-2\nu\sqrt{\rho_{1}\rho_{2}}\left[\cos(\phi_{a})-\frac{\lambda^{2}}{2}\sin^{2}(\phi_{a})\right],
\end{equation}
where $\lambda^{2}=2\rho_{0}\delta^{2}\sqrt{(g_{11}-g_{12})(g_{22}-g_{12})}/\nu$.
The potential is flattened around $\phi_{a}=\pi$ at $\lambda^{2}\lesssim1$
and develops a local minimum at $\lambda^{2}>1$. The former is relevant
for the studies of the slow-roll of a scalar field, while the later
is relevant for the studies of the quantum decay of a scalar field
from a metastable minimum. These two fundamental scenarios can be
both realized in our system. We have shown that the potential energy
$U(\phi_{a})$ in the BEC theory, and the effective scalar field potential
$V(\phi_{a})$ are proportional. Next we will use functional integral
methods to analyze the phase dynamics.

\section{Path-Integral representation\label{sec:Functional-Integral-representati}}

In the following section, to facilitate calculations, we will set
$g_{12}=0$, $g_{11}=g_{22}=g$, $\rho_{1}=\rho_{2}=\rho_{0}$, so
that $\lambda^{2}$ is given in Eq.~\eqref{eq:Lamnda}.

We introduce the quantum partition function~\cite{Atland2010}
\begin{equation}
{\cal Z}=\int{\cal D}(\psi^{\ast},\psi){\rm e^{-S[\psi^{\ast},\psi]}},
\end{equation}
where $S[\psi^{\ast},\psi]=\int d{\bf s}\left[\psi_{j}^{\ast}\partial_{\tau}\psi_{j}+H_{{\rm eff}}(\psi^{\ast},\psi)\right]$
is the action. Here, ${\bf s}=(\tau,{\bf r})$ is a $1+d$ vector,
where $\tau=it/\hbar\in[0,\beta]$ is imaginary time. The quantum
path integral is taken over all configurations of the complex field
$\psi_{j}(\tau,{\bf r})$ with the periodic boundary condition $\psi_{j}({\rm {\bf \beta},{\bf r}})=\psi_{j}(0,{\bf r})$.

We look first for a static solution to identify vacua. This amounts
to replacing $\psi_{j}=\psi_{0}={\rm const}$ in the saddle-point
approximation $\delta S/\delta\psi_{j}=0$. For $\nu>0$ we obtain
the stable $|\psi_{0}|^{2}=(\mu+\nu)/g(1-\delta^{2})$ and unstable
$|\psi_{0}|^{2}=(\mu-\nu)/g(1-\delta^{2})$ vacua. These correspond
to the two Bose gases being in phase and out-of-phase respectively,
although this depends on the sign chosen for the microwave coupling
term $\nu$.

Let us introduce new field variables with the definitions $\psi_{j}({\bf s})=\rho_{j}^{1/2}({\bf s})e^{i\phi_{j}({\bf s})}$,
where $\rho_{j}({\bf s})=\rho_{0}+\delta\rho_{j}({\bf s})$ and $\rho_{0}=|\psi_{0}|^{2}$.
The variables $\delta\rho$ and $\phi$ parametrize the deviation
of the Bose fields from a vacuum. Substituting this parametrization
of the fields into the action, we obtain

\begin{equation}
\begin{split}S & \approx\int d{\bf s}\left\{ i\delta\rho_{j}\partial_{\tau}\phi_{j}+\frac{\hbar^{2}\rho_{0}}{2m}(\nabla\phi_{j})^{2}+\frac{\hbar^{2}}{2g}V(\phi_{a})\right.\\
 & +\frac{[2g\rho_{0}(1-\delta^{2})+\nu\cos(\phi_{a})]\delta\rho_{j}^{2}}{4\rho_{0}}+\frac{\hbar^{2}(\nabla\delta\rho_{j})^{2}}{8m\rho_{0}}\\
 & -\nu\left[\cos(\phi_{a})-\lambda^{2}\sin^{2}(\phi_{a})\right]\delta\rho_{j}\\
 & \left.-\frac{\nu}{4\rho_{0}}\left[\cos(\phi_{a})-2\lambda^{2}\sin^{2}(\phi_{a})-\lambda^{2}\right]\delta\rho_{j}\delta\rho_{3-j}\right\} .
\end{split}
\label{eq:S_rho_fi}
\end{equation}
Here $\phi_{a}=\phi_{1}-\phi_{2}$ is the relative phase and $V(\phi_{a})$
is the effective potential given in Eq.~\eqref{eq:Potential}.

The potential $V(\phi_{a})$ develops a local minimum at $\phi_{a}=\pm\pi$
for $\lambda^{2}>1$ (cf. Fig.~\ref{fig:Effective-field-potential}),
which corresponds to a false vacuum. Multiple equivalent true vacua
occur at the global minima with $\phi_{0}=\pm0,2\pi,4\pi,\ldots$

Since the action is now quadratic in the density fields, we can perform
a Gaussian integration over the density fields. Ignoring gradients
acting on the density fields (i.e.,\ the terms $\hbar^{2}(\nabla\delta\rho_{j})^{2}/8m\rho_{0}$)
in comparison with the potential cost of these fluctuations (i.e.\,,
the terms $g\delta\rho_{j}^{2}/2$) and introducing relative and total
phases, $\phi_{a}=\phi_{1}-\phi_{2}$ and $\phi_{t}=\phi_{1}+\phi_{2}$
respectively, we obtain

\begin{gather}
S\approx\int d{\bf s}\left\{ \frac{(\partial_{\tau}\phi_{a})^{2}}{4g[1+\tilde{\nu}\cos(\phi_{a})-\tilde{\nu}\lambda^{2}]}+\frac{\hbar^{2}\rho_{0}}{4m}(\nabla\phi_{a})^{2}+\frac{\hbar^{2}}{2g}V'(\phi_{a})\right.\nonumber \\
\left.+\frac{1}{4g}(\partial_{\tau}\phi_{t})^{2}+\frac{\hbar^{2}\rho_{0}}{4m}(\nabla\phi_{t})^{2}+\frac{\nu}{g}i\partial_{\tau}\phi_{t}F[\phi_{a}]\right\} .\label{eq:ActionPhases}
\end{gather}

Here we have denoted

\begin{equation}
F[\phi_{a}]=\cos(\phi_{a})-\lambda^{2}\sin^{2}(\phi_{a}),
\end{equation}
and introduced an effective potential for the relative phase modified
by the presence of the environment as

\begin{equation}
V'(\phi_{a})=V(\phi_{a})+2\frac{\nu^{2}}{\hbar^{2}}\left[\cos(\phi_{a})-\lambda^{2}\sin^{2}(\phi_{a})\right]^{2}.\label{eq:NewPotential}
\end{equation}

The action~\eqref{eq:ActionPhases} contains two fields, the relative
and the total phases, coupled by the last term in Eq.~\eqref{eq:ActionPhases}.
The total phase field is characterized by the speed of sound $c=\sqrt{g\rho_{0}/m}$.
In contrast to this, the relative phase is characterized by a modified
speed of sound, which in the semiclassical limit $\lambda\gg1$ (see
below) is given by $c_{a}=\sqrt{g\rho_{0}(1-\tilde{\nu}\lambda^{2})/m}$,
where $\tilde{\nu}=\nu/(g\rho_{0})$.

Apart from some corrections from the environmental degrees of freedom,
which are higher order effects that are omitted here for simplicity,
the effective action for the relative phase now reads

\begin{equation}
S_{a}(\phi_{a})\approx\frac{\hbar^{2}}{2g}\int d{\bf s}\left[\frac{1}{2\hbar^{2}}(\partial_{\tau}\phi_{a})^{2}+\frac{c_{a}^{2}}{2}(\nabla\phi_{a})^{2}+V'(\phi_{a})\right].\label{eq:Sa}
\end{equation}
This action corresponds to the equation of motion given in Eq.~\eqref{eq:Hubble-1}
with the replacement $\phi\rightarrow\phi_{a}$ and $V\rightarrow V'(\phi)$
given in Eq.~\eqref{eq:NewPotential}. We have thus arrived at a
quantum field model similar to Coleman's original model for the false
vacuum decay.

The action in Eq.~\eqref{eq:ActionPhases} is quadratic in the total
phase fields $\phi_{t}$ and we can again perform a Gaussian integration,
this time over the total phase fields. This yields our final result,
that includes the effect of the environmental density fluctuations
omitted in Eq. (\ref{eq:Sa}):

\begin{equation}
\begin{split}S(\phi_{a})= & \frac{\hbar^{2}}{2g}\int d{\bf s}\left[\frac{1}{2\hbar^{2}}(\partial_{\tau}\phi_{a})^{2}+\frac{c_{a}^{2}}{2}(\nabla\phi_{a})^{2}+V'(\phi_{a})\right]\\
 & +\frac{\nu^{2}}{g}\int d{\bf s}\int d{\bf s}^{\prime}F[\phi_{a}({\bf s})]F[\phi_{a}({\bf s}^{\prime})]{\cal G}({\bf s}-{\bf s}^{\prime}),
\end{split}
\label{eq:action_with_damping}
\end{equation}
where
\begin{equation}
\begin{split}{\cal G}(\tau,x) & =\frac{1}{\beta L^{d}}\sum_{\omega_{n},{\bf k}}e^{-i(\omega_{n}\tau+kx)}\frac{\omega_{n}^{2}}{\omega_{n}^{2}+(c\hbar k)^{2}}\end{split}
\end{equation}
is a non-local kernel responsible for long-range correlations in the
relative phase sector induced by the environmental degrees of freedom.
For example, at zero temperature and one spatial dimension it can
be evaluated as

\begin{equation}
{\cal G}(\tau,x)=\frac{\hbar c}{2\pi}\frac{x^{2}-(\hbar c)^{2}\tau^{2}}{[x^{2}+(\hbar c)^{2}\tau^{2}]^{2}}.\label{eq:Kernel1D}
\end{equation}
In two and three dimensions expressions for the corresponding higher
dimensional kernel ${\cal G}(\tau,{\bf r})$ are more involved, and
will not be treated here. The effect of the additional non-local term
(the second line in Eq.~\eqref{eq:action_with_damping}) on the dynamics
of the relative phase are explored in the next section.

\section{Tunneling rate\label{sec:Tunneling-rate}}

To quantify the tunneling process, we calculate the probability that
the system has not yet decayed at time $t$. At long time scales it
should behave as ${\cal F}=\exp(-\Gamma t)$~\cite{Takagi2006},
where $\Gamma$ is the decay rate from the false vacuum. In the weak
tunneling limit it can be written in the form $\Gamma=A\exp(-B)$.
The coefficients $A\propto B^{2}$ and $B$ were calculated in the
context of the false vacuum decay using the instanton technique in
Refs.~\cite{Coleman1977,Callan1977}, in some limiting cases. The
decay rate of phase slips in the O(2) quantum rotor model was calculated
in Ref.~\cite{Danshita12}.

Here we use the instanton technique to estimate the coefficient $B$
and the form of the bubbles. This approach is not strictly valid for
shallow potentials with $\lambda\rightarrow1$, but it should provide
reasonable estimates provided that $\lambda-1$ is not too small.
Calculating the coefficitient $B$ allows to extract the decay rate
$\Gamma$ to a level of exponential accuracy. This estimate for $\Gamma$
is independent on the initial state. Quantum corrections are hidden
in the coefficient $A$. However, the calculation of the coefficient
$A$ is a rather complicated problem in quantum field theory~\cite{Callan1977}
and will be omitted here. We focus in particular on certain scaling
relations which appear universal. These will be verified in detailed
numerical quantum dynamical simulations in the next section.

In order to calculate $B$ we need to find first a bounce solution
of the equation of motion corresponding to the imaginary-time action~\eqref{eq:Sa}.
We treat the induced additional term in Eq.~\eqref{eq:action_with_damping}
as perturbation to be included later. Varying the action~\eqref{eq:Sa}
with respect to the field $\phi_{a}({\bf s})$ and setting $R=\sqrt{{\bf r}^{2}+(c_{a}\hbar\tau)^{2}}$,
we find a solution $\phi_{B}$ to the equation of motion, given by

\begin{equation}
\left(\partial_{R}^{2}+\frac{d}{R}\partial_{R}\right)\phi_{B}=c_{a}^{-2}\partial_{\phi_{B}}V'(\phi_{B}),\label{eq:Bounce}
\end{equation}
which must be solved subject to the boundary condition $\phi_{B}(R=\infty)=\pi$
and $\partial_{R}\phi_{B}(R=0)=0$~\cite{Coleman1977}.

Here, $d$ is the total number of space dimensions, since this equation
is also valid in higher dimensions. Eq.~\eqref{eq:Bounce} describes
a fictitious particle with coordinate $\phi_{B}$. It is released
at rest at some position $\phi_{B}(R=0)$ and approaches $\phi_{B}(R=\infty)=\pi$
at long times. For not too small $\lambda-1$, we may assume $\phi_{B}(R=0)=0$
or $\phi_{B}(R=0)=2\pi$ and adopt the thin-wall approximation by
ignoring the friction term in Eq.~\eqref{eq:Bounce}.

The bounce solution can now be easily found from Eq.~\eqref{eq:Bounce}

\begin{equation}
\phi_{B}(R)\approx2\arctan\left(\exp\left[\frac{\lambda\omega_{0}(R-R_{B})}{c_{a}}\right]\right),\label{eq:Anzatz}
\end{equation}
where $R_{B}$ is the radius of the bubble. Inside the bubble ($R\ll R_{B}$)
we get the true vacuum solution of $\phi_{B}(R)=0$ or $2\pi$, while
outside the bubble ($R\gg R_{B}$) we get $\phi_{B}(R)=\pi$ as expected.

The coefficient $B$ can now be calculated as $B=S_{a}[\phi_{B}]$,
or more explicitly

\begin{equation}
B=\Omega_{d+1}\frac{\hbar c_{a}}{2g}\int_{0}^{\infty}dRR^{d}\left[\frac{1}{2}(\partial_{R}\phi_{B})^{2}+\frac{1}{c_{a}^{2}}V'(\phi_{B})\right],\label{eq:B0}
\end{equation}
where $\Omega_{d+1}=2\pi^{(d+1)/2}/\Gamma[(d+1)/2]$ is the solid
angle in $d+1$ total space-time dimensions. Using the bounce solution~\eqref{eq:Anzatz}
we get the semiclassical estimate $B=\Omega_{d+1}R_{B}^{d}\lambda\hbar\omega_{0}\left(1-R_{B}\omega_{0}/2\lambda c_{a}\right)/g$.

Minimizing $B$ with respect to $R_{B}$ we obtain the radius of the
nucleated bubble $R_{B}=2d\lambda c_{a}/[\omega_{0}(d+1)]$ and

\begin{equation}
B=\Omega_{d+1}\frac{\lambda\hbar\omega_{0}}{dg}\left[\frac{2d\lambda c_{a}}{\omega_{0}(d+1)}\right]^{d}.\label{eq:B}
\end{equation}
Once a bubble with radius $R_{B}$ and rate $\Gamma\propto\exp(-B)$
is nucleated, it expands with the speed $c_{a}$, which is slightly
smaller than the speed of sound.

In one dimension we get $B=2\pi\lambda^{2}\hbar c_{a}/g\propto\lambda^{2}\rho_{0}\xi$,
where $\xi=\hbar/\sqrt{2mg\rho_{0}}$ is the healing length. We now
substitute $\phi_{B}(R)$ into the second line of Eq.~\eqref{eq:action_with_damping}.
After a proper rescaling of variables under the integral this leads
to the correction $\Delta B\approx\lambda^{4}\tilde{\nu}^{2}\rho_{0}\xi$.
Combining with the semiclassical estimate we arrive at

\begin{equation}
B_{1D}=\left(\beta(\lambda)+\gamma(\lambda)\tilde{\nu}^{2}\right)\rho_{0}\xi.\label{eq:B1D}
\end{equation}
Here $\beta(\lambda)$ and $\gamma(\lambda)$ are complicated expressions
in general, and cannot be easily obtained precisely outside of the
thin-wall limit of Coleman.

We note that the theory generally predicts a quadratic dependence
on $\tilde{\nu}^{2}$, which we show does agree with quantitative
numerical simulations in the next section. At large values of $\lambda$
the semiclassical analysis in the thin-wall limit yields $\beta(\lambda)\propto\lambda^{2}$and
$\gamma(\lambda)\propto\lambda^{4}$. It also follows that the effect
of the non-local correlations on the quantum tunneling process in
one dimension is small in the sine-Gordon regime where $\tilde{\nu}\ll1$.

\section{Numerical Analysis\label{sec:Numerical-Analysis} }

The path integral calculations given above are indicative of the potential
for simulating the decay of a relativistic quantum field metastable
vacuum using an ultra-cold atomic BEC.

Yet how practical is this, really? How accurate are the approximations
used? Most crucially, how long will tunneling take? This last question
is an important one, because current laboratory BEC experiments are
limited in time duration by trap losses. These in turn depend on many
issues, ranging from the vacuum quality to the size of nonlinear loss
effects due to collisions. Depending on the isotope used and the density,
the lifetime typically varies between millisecond to seconds, in current
experiments.

Neither the tunneling prefactor $A$ nor the exponent $B$ is easily
calculable for our system. Even B is known only in the simplest of
cases, so it is not possible to analytically obtain an estimated tunneling
time. We instead resort to numerical simulations of the full quantum
field dynamics, which has a number of advantages. The full dynamics
of using BECs with modulated coupling is easily included, and one
can also include laboratory losses. Most significantly, one is not
restricted to the slow tunneling, deep well regime as in Coleman's
original work. This is fortunate, since the slow tunneling regime
is neither suited to experiments, nor well matched to currently proposed
cosmological models.

The truncated Wigner approximation (TWA), where a quantum state is
represented by a phase space distribution of stochastic trajectories
following the Gross-Pitaevskii equation~\cite{Drummond1993a,Steel1998}
together with dissipative noise terms, enables one to capture many
quantum features of the system. This method gives the first quantum
correction to the Gross-Pitaevskii equation, in an expansion in $M/N$,
where $M$ is the number of modes and $N$ is the total number of
bosons. It is known to correctly predict quantum fluctuation dynamics
in a number of quantitative experiments at the quantum noise level~\cite{Drummond2016}.
We use this approach to perform stochastic numerical simulations on
the full BEC model in order to investigate true vacuum nucleation
numerically. The TWA generally needs to be checked with more precise
methods~\cite{Deuar:2006}. From previous work we expect it to be
able to generate tunneling times that are accurate enough to give
estimates of useful experimental parameter values.

This approach was originally used to predict quantum squeezing dynamics
in photonic quantum solitons~\cite{Carter1987a,Drummond1993b}, which
have a combination of quantum field propagation and dissipative coupling
to phonon reservoirs. The truncation approximation was justified by
comparisons to exact positive-P quantum dynamical theory~\cite{Drummond1993a},
and gave excellent quantitative, first principles agreement with high-precision
experimental measurements~\cite{Corney2006c}. Succesful comparisons
with BEC interferometry in three dimensions have been made, showing
that this approach is applicable to bosonic atoms. Tunneling in shallow
potentials for related parametric systems~\cite{Drummond1989} has
been treated, giving agreement with exact methods. However, the truncated
Wigner method can become inaccurate when treating deeper wells, or
systems where scattering into unoccupied vacuum modes in three dimensions
is a dominant feature~\cite{Deuar:2007_BECCollisions}. This is a
possible limitation when treating tunneling in higher space dimensions,
since the true vacuum spatial modes will not be highly occupied initially.

We simulate the false vacuum decay in one spatial dimension. BEC is
impossible in low-dimensional homogenous systems, but it should occur
when atoms are trapped because the confining potential modifies the
density of states~\cite{Bagnato91}. It can be realized in low-dimensional
optical and magnetic traps in which the energy-level spacing exceeds
interparticle interaction and temperature~\cite{Ketterle01}, and
we give a quantitative estimate of the relevant coherence length in
the last section. The initial quantum state is assumed to be a highly
occupied Bose-Einstein condensate in a coherent state. The coherent
state distribution for the corresponding Wigner phase-space distribution
is a Gaussian in phase space, and the vacuum modes have a corresponding
initial variance of $n=1/2$. Our initial state construction for the
Wigner representation then proceeds by simply including vacuum noise
for each mode to the coherent state. The primary physical effect of
the initial noise is to allow spontaneous scattering processes that
are disallowed in pure Gross-Pitaevskii theory. Additional noise is
required when there is damping, in order to correctly model the fluctuation-dissipation
properties of a quantum phase-space representation of this type.

Typical results are presented as single trajectories in the plots
included here, for illustrative purposes. Given that the universe
is, in current cosmology, a single quantum wavefunction, it is a subtle
question to understand how these trajectories can be compared to cosmological
events. The use of a Wigner representation to generate such trajectories
is motivated by the fact that the marginal probabilities of the Wigner
distribution are true, classical-like probabilities~\cite{Hillery_Review_1984_DistributionFunctions}.
This depends on the observations~\cite{lewis2016approximate}, and
is surely only a first step towards understanding this hypothesis.

As the predicted behaviour is stochastic, the computed tunneling rates
to be compared with a future experiment require and utilize a full
quantum ensemble average. More fundamentally, our philosophy is that
one should understand the proposed experiment as a quantum computer
for this unsolved dynamical quantum field problem. Given that our
computer simulations embody a truncation approximation, they should
not be regarded as definitive. An indicator of the breakdown of the
TWA would be tunneling rates that are faster than anticipated.

The chief limitation of this approach is that it neglects one of the
likely features of an experiment, which is that the initial state
will be generated dynamically by Rabi rotating a finite temperature
equilibrium ensemble, which could alter the tunneling times. We do
not treat this in the present study. However, the use of an initial
coherent state does correctly model the correlations induced by the
Rabi rotation process. Methods that treat finite temperature effects
are known~\cite{Ruostekoski2005}, and will be carried out in another
publication. While it is certainly possible that the precise initial
quantum state may have an effect on tunneling, the best candidate
quantum state for early universe modeling is not well understood.
One longer term goal of this research is to determine whether cosmological
data can throw any light on this question, which will require further
modeling and experiments with different initial quantum states.

\subsection{Symmetric BEC experiment\label{subsec:Symmetric-BEC-experiment}}

The first case we consider is the idealized case treated analytically
in the previous section, with equal intra-state scattering lengths,
and zero inter-state scattering lengths. The stationary solution of
the two independent condensates at the classical level is found by
solving the Gross-Pitaevskii equation in imaginary time, without any
linear coupling between the two species. The initial conditions are
$\psi_{j}=\sqrt{\rho_{0}}\exp[i(j-1)\pi]$, such that the number densities
$\rho_{j}=|\psi_{j}|^{2}$ are the same and the relative phase is
$\pi$. Next, quantum noise corresponding to a coherent state is added
to this state as $\psi(x)\rightarrow\psi(x)+\sum_{i=1}^{M}\alpha_{i}\exp(ik_{i}x)/\sqrt{L}$.
Here $\alpha_{i}$ are complex Gaussian variables with $\overline{\alpha_{i}^{\ast}\alpha_{k}}=\delta_{ik}/2$,
thus sampling vacuum or coherent state fluctuations\@. The number
of modes $M$ is chosen to represent the physical system, while being
much smaller than the total number of atoms $N$, so that $M/N\ll1$,
as required for the truncated Wigner method.

We propagate this state in real time by solving the time-dependent
coupled mean-field equations~\eqref{eq:GP}. In the absence of damping
and noise, the corresponding equations for the Wigner representation
are:

\begin{equation}
i\hbar\frac{d\Psi_{j}}{dt}=-\frac{\hbar^{2}}{2m}\frac{\partial^{2}\Psi_{j}}{\partial x^{2}}+g\Psi_{j}\left(|\Psi_{j}|^{2}-\delta_{M}-\frac{\mu}{g}\right)-\nu_{t}\Psi_{3-j}.\label{eq:GP_Sym}
\end{equation}
where for the plane wave basis $\delta_{M}=M/L$. For our numerical
calculations we use dimensionless variables. These are defined by
scaling time in terms of the characteristic frequency $\omega_{0}=2\sqrt{\nu g\rho_{0}}/\hbar$,
and scaling space by combining this with the speed of sound $c=\sqrt{g\rho_{0}/m}$.
The resulting units of time are $\tilde{t}=t\omega_{0}$, length $\tilde{x}=x/x_{0}=x\omega_{0}/c$
and energy $\tilde{\nu}=\nu/g\rho_{0}$.

The equations of motion in the new units become

\begin{gather}
i\frac{d\tilde{\Psi}_{j}}{d\tilde{t}}=\left[-\sqrt{\tilde{\nu}}\frac{\partial^{2}}{\partial\tilde{x}^{2}}+\frac{|\tilde{\Psi}_{j}|^{2}-\tilde{\delta}_{M}}{2\sqrt{\tilde{\nu}}\tilde{\rho}_{0}}+\frac{\tilde{\nu}-1}{2\sqrt{\tilde{\nu}}}\right]\tilde{\Psi}_{j}\nonumber \\
-\frac{\sqrt{\tilde{\nu}}}{2}\left[1+\sqrt{2}\lambda\tilde{\omega}\cos(\tilde{\omega}\tilde{t})\right]\tilde{\Psi},\label{eq:GP_Sym2}
\end{gather}
where $\lambda=\delta\sqrt{2}/\sqrt{\tilde{\nu}}$, $\tilde{\rho}_{0}=\rho_{0}x_{0}$
and $\tilde{\delta}_{M}=Mx_{0}/L$.

\begin{figure}[!t]
\includegraphics{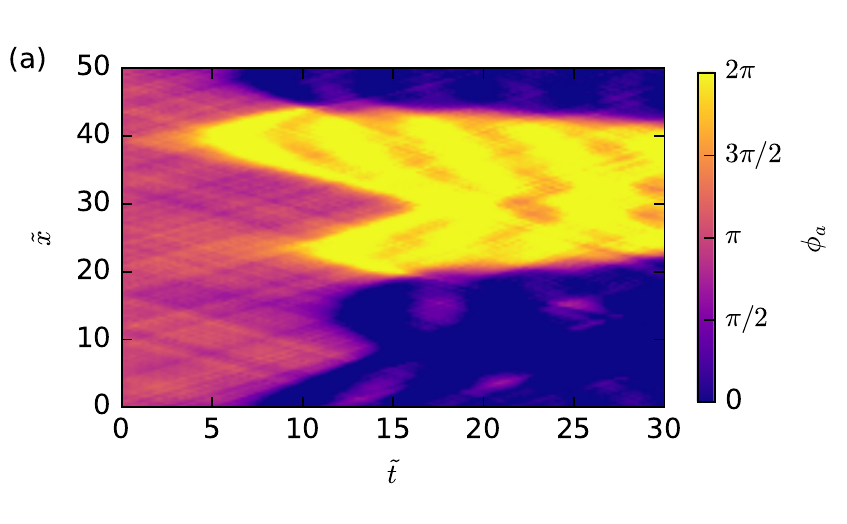}\\
\includegraphics{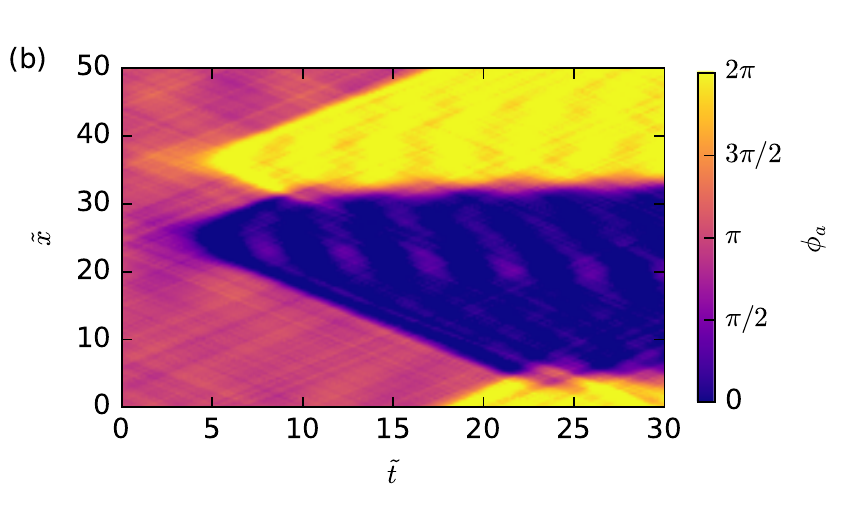}

\caption{\textbf{\label{fig:1d-bubbles-1}}Decay of the false vacuum in 1D.
(a) A single-trajectory simulation of the false vacuum decay in 1D
with $N_{\mathrm{grid}}=256$ and dimensionless parameters $\lambda=1.3$,
$\tilde{\omega}=50$, $\tilde{\nu}=2\times10^{-3}$, $\tilde{L}=50$,
$\tilde{\rho}_{0}=1000$, $a_{11}=3a_{0}$, $a_{22}=20a_{0}$, $a_{12}=0$
(corresponding to a two-component $^{7}$Li condensate near $640\,\mathrm{G}$
resonance in a ring trap with $N=5\times10^{4}$, trap circumference
$L=121.7\,\mathrm{\mu m}$, transverse frequency $\omega_{\perp}=2\pi\times61\,\mathrm{kHz}$,
observation time $T=1.75\,\mathrm{ms}$, oscillator amplitude $\nu/\hbar=2\pi\times61\,\mathrm{Hz}$,
frequency $\omega=2\pi\times136.4\,\mathrm{kHz}$ and modulation $\delta=0.041$).
The initial component population ratio $N_{1}/N_{2}\approx0.872/0.128$
(obtained by solving Eq.~\eqref{eq:saddlepointconditions} numerically
for $\theta$, with $N_{1}/N_{2}\equiv\cot^{2}\theta$). (b) Same
parameters as in the previous panel, except the scattering lengths
have been set to $a_{11}^{\prime}=a_{22}^{\prime}=\left(a_{11}+a_{22}\right)/2$,
to give the symmetric case.\label{fig:1D-1}}
\end{figure}

The results of typical simulations in one dimension are shown in Fig.~\ref{fig:1d-bubbles-1}.
The single trajectory dynamics features the creation of four bubbles
in the asymmetric case, treated in detail in the next section, or
two bubbles in the symmetric case.

We note here a highly characteristic feature of this type of tunneling,
which is the creation of topologically distinct vacua depending on
whether the tunneling occurred with a positive or negative phase change.
Even though these vacua are locally indistinguishable, they are globally
distinct. In the absence of large density changes, they cannot combine
with each other, as they are necessarily separated by a high energy
region of false vacuum. Collisions of bubbles result either in the
creation of localized oscillating structures known as oscillons~\cite{Amin2012},
or domain walls if the colliding bubbles belong to topologically distinct
vacua. Such topologically distinct vacua, if they occurred in the
real universe, would presumably have to occur at cosmologically large
separations.

In our numerical simulations of tunneling, we fix $\tilde{\rho}_{0}=200$.
Therefore, recalling that $\tilde{\nu}=\nu/g\rho_{0}$, we can express
the density dependent factors in Eq.~\eqref{eq:B1D} in terms of
the scaled coupling constant $\tilde{\nu}$ as:
\begin{equation}
\rho_{0}\xi=\tilde{\rho}_{0}\xi/x_{0}=\tilde{\rho}_{0}\omega_{0}\xi/c=\tilde{\rho}_{0}\sqrt{2\tilde{\nu}}\,.
\end{equation}
Substituting this expression for $\rho_{0}\xi$ into Eq.~\eqref{eq:B1D}
we obtain an expected scaling law at fixed $\tilde{\rho}_{0}$ where
the exponents increase by $1/2$, so that:

\begin{equation}
B_{1D}(\lambda,\tilde{\nu})=\tilde{\beta}(\lambda)\tilde{\nu}^{1/2}+\tilde{\gamma}(\lambda)\tilde{\nu}^{5/2},\label{eq:Brescaled}
\end{equation}
where $\tilde{\beta}=\sqrt{2}\tilde{\rho}_{0}\beta$ and $\tilde{\gamma}=\sqrt{2}\tilde{\rho}_{0}\gamma$.

\begin{figure}[!t]
\includegraphics[width=1\columnwidth]{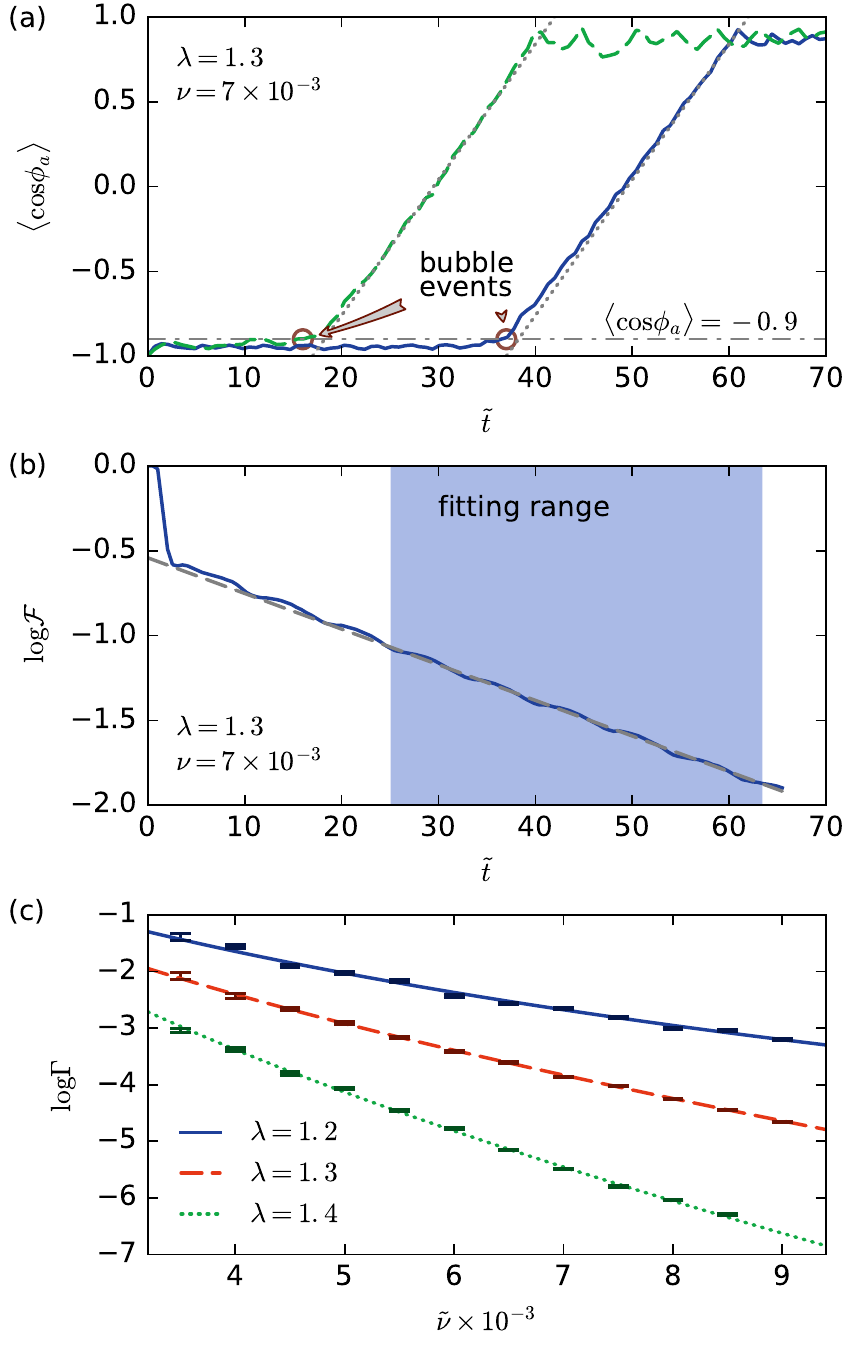}

\caption{(a) Average relative phase $\langle\cos\phi_{a}\rangle$ in two separate
single-trajectory simulations. A bubble is detected at $\tilde{t}=16$
and $\tilde{t}=37$. (b) Logarithmic plot of the survival probability
$\mathcal{F}$ estimated using $16384$ TWA trajectories. The standard
error in the mean is smaller than the line thickness and, therefore,
not shown in the graph. The long time scale part of the plot (marked
with the blue background) is fitted with a linear dependence (that
is, ${\cal F}=\exp(-\Gamma\tilde{t})$), the slope of which gives
the negated tunneling rate $-\Gamma$. (c) Dependence of the tunneling
rate $\Gamma$ on the coupling $\tilde{\nu}$ for different values
of $\lambda$. The errorbars show the combination of two sources of
errors: the standard error in the mean of $\mathcal{F}$ and the estimated
error of the linear fitting of $\log\mathcal{F}$. We fitted the data
with $\log(\Gamma)=\tilde{\alpha}+\tilde{\beta}\tilde{\nu}^{1/2}+\tilde{\gamma}\tilde{\nu}^{5/2}$,
where $\{\tilde{\alpha},\tilde{\beta},\tilde{\gamma}\}=\{1.72(8),-1.69(3),6.2(4)\times10^{-4}\}$
for $\lambda=1.2$, $\{1.9(2),-2.14(9),-4(3)\times10^{-4}\}$ for
$\lambda=1.3$, $\{\tilde{\alpha},\tilde{\beta},\tilde{\gamma}\}=\{2.9(4),-3.1(2),-6(11)\times10^{-4}\}$
for $\lambda=1.4$.\label{fig:TunnelingRate}}
\end{figure}

Varying $\tilde{\nu}$ and keeping $\lambda$ fixed allows us to study
the quantum dynamics of the scalar field. Our TWA approach is expected
to yield accurate predictions for the relatively shallow effective
potentials necessary for tunneling over laboratory time-scales~\cite{Drummond1989}.
In Fig.~\ref{fig:TunnelingRate} we present the scaling of the tunneling
rate of the bubbles as a function of $\tilde{\nu}$ for different
values of $\lambda$. The fitting procedure is as follows.

In each trajectory we define the appearance of a bubble as the average
relative phase $\langle\cos\phi_{a}\rangle\equiv\frac{1}{\tilde{L}}\int_{0}^{\tilde{L}}\cos\phi_{a}(\tilde{x})d\tilde{x}$
first exceeding the (empyrically chosen) threshold of $0.9$, as illustrated
in Fig.~\ref{fig:TunnelingRate}(a).

From the set of times $\tilde{t}_{j}^{\mathrm{bubble}}$ obtained
this way in each of the $N_{\mathrm{tr}}$ TWA trajectories we approximate
the decay probability as $\mathcal{F}_{\mathrm{decay}}\left(\tilde{t}\right)\equiv\int_{0}^{t}{\cal P}(\tilde{t}^{\prime})d\tilde{t}^{\prime}\approx\left|\left\{ j,\,\tilde{t}_{j}^{\mathrm{bubble}}<\tilde{t}\right\} \right|$.
At each point of time we can consider the set of decayed/non-decayed
trajectories to be a sample from a binomial distribution and estimate
the standard error of the mean as $E_{\mathcal{F}}\approx\sigma\left(\mathcal{F}\right)/\sqrt{N_{\mathrm{tr}}}=\mathcal{F}_{\mathrm{decay}}\left(1-\mathcal{F}_{\mathrm{decay}}\right)\sqrt{N_{\mathrm{tr}}}$.
The ``survival'' probability defined as ${\cal F}\left(\tilde{t}\right)=1-\mathcal{F}_{\mathrm{decay}}$
for one of the values of $\lambda$ and $\tilde{\nu}$ is plotted
in Fig.~\ref{fig:TunnelingRate}(b). The propagated error of $\log\mathcal{F}$,
$E_{\mathcal{F}}/\mathcal{F_{\mathrm{decay}}}$, is smaller than the
thickness of the line and is not shown on the plot.

The long time scale region is defined as the part of the evolution
between the values of the relative population difference $P_{z}=0$
and $P_{z}=-0.5$. The relative population difference is calculated
as $P_{z}=\left(N_{2}-N_{1}\right)/\left(N_{1}+N_{2}\right)$, where
$N_{1}$ and $N_{2}$ are the component populations after a $\pi/2$
Rabi rotation. The ``survival'' probability in this region is fitted
with an exponential, shown as the dashed grey line in Fig.~\ref{fig:TunnelingRate}(b)
using weighted least squares (WLS), with weights being the errors
of the mean from the previous step. This resultes in the value of
the slope of the log plot $-\Gamma$, along with the associated error,
for each pair of $\tilde{\nu},\lambda$. These points with errorbars
are shown in Fig.~\ref{fig:TunnelingRate}(c).

Since $\Gamma=A\exp(-B)$, where $B$ is given in Eq.~\eqref{eq:Brescaled},
we fit the curves with $\log(\Gamma)=\tilde{\alpha}+\tilde{\beta}\tilde{\nu}^{1/2}+\tilde{\gamma}\tilde{\nu}^{5/2}$,
where $\tilde{\alpha}=\log(A)$. We were only able to fit the value
of $\tilde{\gamma}$ accurately for $\lambda=1.2$, while for higher
values of $\lambda$ the estimated error is comparable to the fitted
value of the coefficient. Overall, the observed behavior provides
strong evidence of a quantum tunneling process leading to bubble nucleation,
with a scaling dependence on $\tilde{\nu}$ that agrees with the path
integral result, Eq.~\eqref{eq:B1D}.

\subsection{Non-zero inter-component interaction\label{subsec:Non-zero-inter-component-interac}}

While the previous analysis treated the symmetric case, not all atomic
species have this type of Hamiltonian. We therefore turn to the more
general case typical, for example, of the broad Feshbach resonance
known to occur in $^{7}$Li.

The most general equation of motion for TWA approach corresponding
to the Hamiltonian~\eqref{eq:Hamiltonian} reads:

\begin{multline}
i\hbar\frac{d\Psi_{j}}{dt}=-\frac{\hbar^{2}}{2m}\frac{\partial^{2}\Psi_{j}}{\partial x^{2}}+\\
\sum_{k=1}^{2}g_{jk}\left(|\Psi_{k}|^{2}-\frac{\delta_{jk}+1}{2}\delta_{M}-\mu\right)\Psi_{j}-\nu_{t}\Psi_{3-j}.
\end{multline}
This equation is similar to Eq.~\eqref{eq:GP_Sym}. We now allow
the inter-component interaction strength $g_{12}\ne0$ as well as
the possibility $g_{11}\ne g_{22}$.

Similarly to the derivation of Eq.~\eqref{eq:GP_Sym2} we introduce
units of time $\tilde{t}=t\omega_{0}$ and length $\tilde{x}=x/x_{0}=x\omega_{0}/c$,
where now $c=\sqrt{\bar{g}\rho_{0}/m}$, $\bar{g}=(g_{11}+g_{22})/2+g_{12}$,
and $\omega_{0}=2\sqrt{\nu\bar{g}\rho_{0}}/\hbar$. The chemical potential
is taken so that $\mu=\bar{g}\rho_{0}+\nu$. Denoting $\tilde{\nu}=\nu/\bar{g}\rho_{0}$,
$\tilde{g}_{jk}=g_{jk}/\bar{g}$, and $\lambda=\delta\hbar\omega_{0}/\sqrt{2}\nu$,
we arrive at

\begin{gather}
i\frac{d\tilde{\Psi}_{j}}{d\tilde{t}}=\left[-\sqrt{\tilde{\nu}}\frac{\partial^{2}}{\partial\tilde{x}^{2}}+\sum_{k=1}^{2}\frac{\tilde{g}_{jk}}{2\sqrt{\tilde{\nu}}\tilde{\rho}_{0}}\left(|\tilde{\Psi}_{k}|^{2}-\frac{\delta_{jk}+1}{2}\tilde{\delta}_{M}\right)\right]\tilde{\Psi}_{j}\nonumber \\
+\frac{\tilde{\nu}-1}{2\sqrt{\tilde{\nu}}}\tilde{\Psi}_{j}-\frac{\sqrt{\tilde{\nu}}}{2}\left[1+\sqrt{2}\lambda\tilde{\omega}\cos(\tilde{\omega}\tilde{t})\right]\tilde{\Psi}_{3-j},\label{eq:Asymmetric}
\end{gather}
where $\tilde{\rho}_{0}=\rho_{0}x_{0}$ and $\tilde{\delta}_{M}=Mx_{0}/L$.

First of all, we need to find appropriate initial conditions for simulating
Eq.~\eqref{eq:Asymmetric}. The dynamics of the classical fields
$\psi_{j}$ is governed by the potential function:

\begin{equation}
{\cal H}=\frac{g_{jk}}{2}|\psi_{k}|^{2}|\psi_{j}|^{2}-\nu\psi_{j}^{\ast}\psi_{3-j},
\end{equation}
where summation over spin indices $j$ and $k$ is assumed as in Eq.~\eqref{eq:Hamiltonian}.

We parameterize the field $\psi_{1}$ and $\psi_{2}$ as

\begin{eqnarray}
\psi_{1} & = & u\exp\left[i\left(\phi_{s}+\phi_{a}\right)/2\right]\cos\theta,\nonumber \\
\psi_{2} & = & u\exp\left[i\left(\phi_{s}-\phi_{a}\right)/2\right]\sin\theta.
\end{eqnarray}
and adopt the dimensional units introduced above to get:

\begin{multline}
{\cal H}\equiv u^{4}g\tilde{{\cal H}}=\\
u^{4}g\left(-\tilde{\nu}\cos\phi_{a}\sin2\theta+\frac{\tilde{g}_{s}}{2}+\frac{\tilde{g}_{sa}}{2}\cos^{2}2\theta+\tilde{g}_{a}\cos2\theta\right).\label{eq:potentialforsaddlepoint}
\end{multline}

The initial conditions are found from the two saddle-point equations
$\partial\tilde{{\cal H}}/\partial\phi_{a}=0$ and $\partial\tilde{{\cal H}}/\partial\theta=0$,
or

\begin{gather}
\tilde{\nu}\sin\phi_{a}\sin2\theta=0,\label{eq:saddlepointconditions}\\
-2\tilde{\nu}\cos\phi_{a}\cos2\theta-2\tilde{g}_{sa}\sin2\theta\cos2\theta-2\tilde{g}_{a}\sin2\theta=0.\nonumber
\end{gather}
Among the various solutions to the above equations, we are interested
in solutions satisfying $\partial^{2}\tilde{V}/\partial\theta^{2}>0$
and

\begin{equation}
\frac{\partial^{2}\tilde{{\cal H}}}{\partial\phi_{a}^{2}}\frac{\partial^{2}\tilde{{\cal H}}}{\partial\theta^{2}}-\left(\frac{\partial^{2}\tilde{{\cal H}}}{\partial\phi_{a}\partial\theta}\right)^{2}<0.\label{eq:saddlepointcondition2}
\end{equation}
In the symmetric case $g_{11}=g_{22}$, we obtain $\psi_{j}=\sqrt{\rho_{0}}\exp[i(j-1)\pi]$.
In the more general case $g_{11}\ne g_{22}$ Eqs.~\eqref{eq:saddlepointconditions}
and Eq.~\eqref{eq:saddlepointcondition2} are solved numerically.
Typical results of these numerical simulations are shown in Fig.~\ref{fig:1d-bubbles-1}.
Although the change in symmetry modifies the dynamics, we see that
the essential feature of false vacuum decay into a true vacuum is
still found.

\subsection{Energy calculations\label{subsec:Energy-calculations}}

\begin{figure}[!t]
\includegraphics{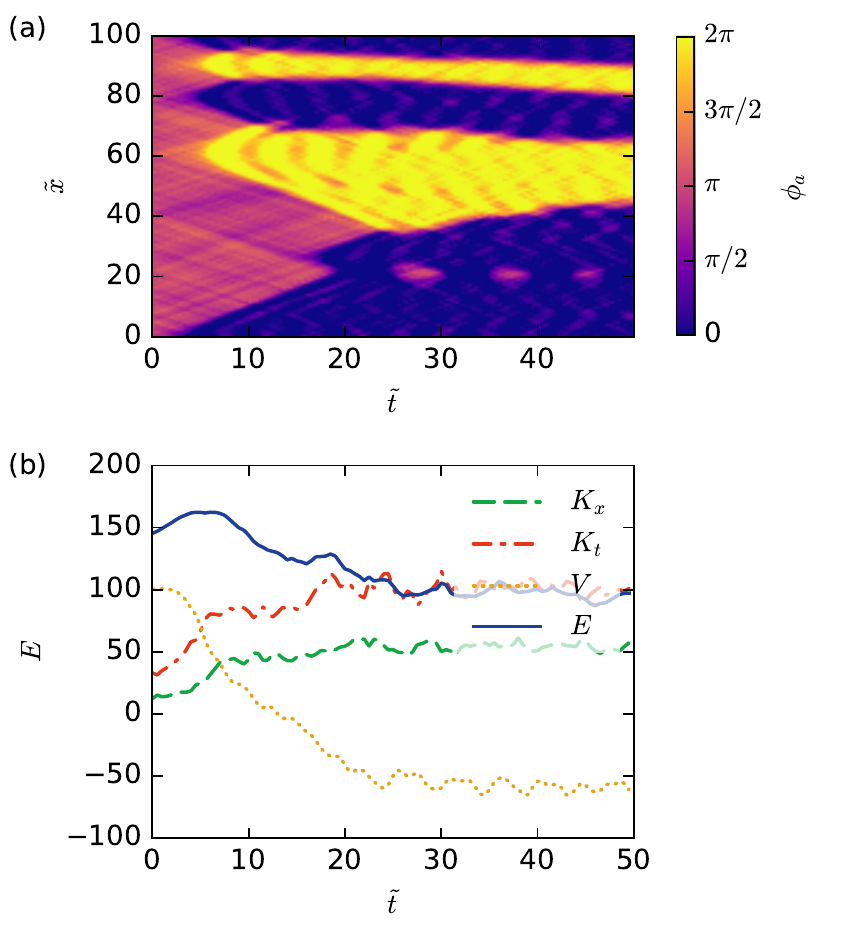}

\caption{\label{fig:energy-1d}Evolution of relative phase (a) and its energy
(b) for $\lambda=1.1$, $\tilde{\omega}=50$, $\tilde{\nu}=0.01$,
$\tilde{\rho}_{0}=200$, $a_{11}=a_{22}$, $a_{12}=0$ with $256$
spatial grid points and $160000$ time steps. }
\end{figure}

We now consider the fate of the potential energy liberated during
vacuum tunneling. In a pure quantum vacuum tunneling theory, all forms
of energy are conserved. Here, it is possible that energy can be converted
from relative phase energy to relative density energy, since density
fluctuations act as a reservoir. This is similar to the way that gravitational
degrees of freedom can modify the quantum field dynamics in inflationary
universe theories.

The energy was calculated as follows. The total energy of the phase
Hamiltonian is:
\begin{equation}
E=V+K_{x}+K_{t},
\end{equation}
 where we divide up the energy into a phase potential energy part,
and contributions from the space and time `kinetic' energy terms,
which correspond to the energy available for free particle creation
in a cosmological interpretation:

\[
V=-\int_{0}^{L}\left(\cos\phi_{a}-\frac{\lambda^{2}}{2}\sin^{2}\phi_{a}\right)d\tilde{x},
\]

\[
K_{x}=\int_{0}^{L}\frac{1}{2}\left(\frac{d\phi_{a}}{d\tilde{x}}\right)^{2}d\tilde{x},
\]

\begin{equation}
K_{t}=\int_{0}^{L}\frac{1}{2}\left(\frac{d\phi_{a}}{d\tilde{t}}\right)^{2}d\tilde{x}\,.
\end{equation}

Fig.~\ref{fig:energy-1d} shows the results of energy calculations
for one TWA trajectory. The energy was measured at $2000$ time points,
then each $20$ values were averaged producing $100$ points in total,
which were plotted. Therefore the averaging time window consisted
of $\tilde{t}_{\max}/\left(2000/20\right)/\left(2\pi/\tilde{\omega}\right)\approx4$
periods of the oscillation of the driving field.

As can be seen from the figure there is strong evidence for conversion
of energy from potential to `kinetic' energy after a tunneling event
has occurred. This implies that, at least at short times after tunneling
has occurred, the phase-sector energy is largely conserved. At longer
times, the total energy stored in the relative phase sector decreases
gradually. We interpret this as transformation of phase-sector energy
to relative density fluctuation energy~\textemdash{} a behaviour
not found in the original Coleman model. This is in accordance with
the path integral results and numerical results presented in Fig.~\ref{fig:TunnelingRate}.
The decrease of the tunneling rate found in these studies is associated
with a damping mechanism in the system, which is clearly seen in the
energy graphs.

\section{Experimental proposal\label{sec:Experimental-proposal}}

The experimental implementation of false vacuum decay in a two-component
Bose-Einstein condensate requires the use of two atomic states with
specific scattering properties: (i) the inter-state scattering length
should be close to zero, and (ii) both states should have positive
scattering lengths so that our parameter $\lambda$ has a real value
(section~\ref{sec:The-model}~A). It is also favorable for clear
observation of this effect to use the states with nearly identical
values of intra-state scattering lengths, in order to avoid the excitation
of collective oscillations and the dynamical evolution of the order
parameter~\cite{Egorov2011}.

Two pairs of Zeeman states in $^{41}{\rm K}$ have these requisite
scattering properties, and can be used to test for the symmetric case
of the false vacuum decay. States $|1\rangle=|F=1,m_{F}=1\rangle$
and $|2\rangle=|F=1,m_{F}=0\rangle$ are predicted~\cite{Lysebo2010}
to have an inter-state Feshbach resonance with a zero-crossing of
the cross-coupling term $a_{12}$ at $675.25\,\mathrm{G}$, and a
resonance width of $0.156\,\mathrm{G}$. The two intra-state scattering
lengths have a similar value: $a_{11}=59.5a_{0}$ and $a_{22}=60.5a_{0}$,
where $a_{0}$ is the Bohr radius. States $|1\rangle$ and $|2\rangle$
are separated by $61.93\,\mathrm{MHz}$ at this magnetic field, and
will be coupled by the amplitude-modulated radio-frequency field via
a magnetic dipole transition. The remaining state $|F=1,m_{F}=-1\rangle$
in this hyperfine manifold is far detuned, with a transition frequency
of $66.1\,\mathrm{MHz}$, and will not participate in the resonant
coupling of the $|1\rangle-|2\rangle$ transition. Another pair of
suitable states consists of states $|1\rangle=|F=1,m_{F}=-1\rangle$
and $|2\rangle=|F=1,m_{F}=0\rangle$, which are also predicted~\cite{Lysebo2010}
to have an inter-state Feshbach resonance with a zero-crossing of
the cross-coupling at $717.6\,\mathrm{G}$. In this case, the resonance
width is $0.118\,\mathrm{G}$, with $a_{11}=61a_{0}$ and $a_{22}=59a_{0}$.

The asymmetric case of the false vacuum decay can be studied with
two Zeeman states $|1\rangle=|F=1,m_{F}=0\rangle$, and $|2\rangle=|F=1,m_{F}=1\rangle$
in a $^{7}$Li condensate. These two states have zero-crossing of
the inter-state scattering length $a_{12}$ at 640 G and positive
values of scattering lengths $a_{11}=3a_{0}$ and $a_{22}=20a_{0}$~\cite{Hulet2015},
as in the conditions of Fig.~\ref{fig:1d-bubbles-1}(a). The $^{7}$Li
condensate can be readily prepared in one of the two Zeeman states,
and the inter-state coupling can be driven by radiofrequency radiation
in a similar way as with $^{41}$K atoms.

We can estimate the phase coherence length for the experimental parameters
quoted in the caption of Fig.~\ref{fig:1d-bubbles-1}(a). The dimensionless
coupling parameter is

\begin{equation}
\gamma=\frac{m\bar{g}}{\hbar^{2}\rho_{0}}=1.25\times10^{-4}.
\end{equation}
The zero-temperature phase coherence length $l_{\phi}^{(0)}\propto e^{2\pi/\sqrt{\gamma}}$
is extremely large. Therefore, the coherence length is limited by
the temperature-dependent phase coherence length $l_{\phi}=\hbar^{2}\rho_{0}/\left(mk_{B}T\right)$,
providing $\tau\equiv T/T_{D}\ll\sqrt{\gamma}$, where the quantum
degeneracy temperature $T_{D}=\hbar^{2}\rho_{0}^{2}/\left(2mk_{B}\right)$~\cite{Bouchoule2012-two-body}.
With our parameters $T_{D}\approx6mK$, so from the condition for
the trap length $L\ll l_{\phi}$ it follows that the restriction on
the temperature is

\begin{equation}
T\ll\frac{\hbar^{2}\rho_{0}}{mk_{B}L}\approx235\,\mathrm{nK}.
\end{equation}

Two Zeeman states can also be coupled by two co-propagating Raman
beams with amplitudes $E_{1}$ and $E_{2}$. In the rotating wave
approximation this results in the Raman coupling between two components
with a strength $\Omega\propto E_{1}E_{2}$~\cite{spielman14}. If
we apply amplitude modulation with frequency $\omega$ to one of the
fields, it is then possible to generate a time-dependent Raman coupling
in the form $\Omega=\Omega_{0}+\text{\ensuremath{\Omega}}_{R}\cos\omega t$,
where the coefficients $\Omega_{0}$ and $\Omega_{R}$ can be controlled
individually~\cite{spielman15}. In this way it is possible to generate
the required Kapitza pendulum coupling with a variable parameter $\lambda$.

The simplest case of a 1D geometry with a uniform distribution of
the atom density along the axial coordinate can be realized in a toroidal
optical dipole trap by the intersection of red-detuned \textquotedblleft sheet\textquotedblright{}
and \textquotedblleft ring\textquotedblright{} laser beams~\cite{PhysRevLett.106.130401}.
One-dimensional evolution of the relative phase can be ensured if
the transverse frequency of the trap ($61\,\mathrm{kHz}$) is larger
than the chemical potential of the condensate. Simulation data shown
in Figs.~\ref{fig:1d-bubbles-1} corresponds to a 1D geometry with
a Bose condensate of $5\times10^{4}$ $^{7}{\rm Li}$ atoms in state
$|1\rangle$, loaded into a ring trap of $39\,\mu m$ diameter. A
$\pi/2$ pulse of resonant r.f.\ radiation prepares a coherent superposition
of states $|1\rangle$ and $|2\rangle$. The Kapitza-pendulum coupling
of two states is realized by an amplitude modulated r.f.\ field phase
shifted by $\pi/2$ in order to prepare the superposition in a metastable
state of the effective potential of Fig.~\ref{fig:Effective-field-potential}.

After some time evolution, an interrogating $\pi/2$ pulse can be
utilized to convert the relative phase distribution along the axial
coordinate into number density distributions $\rho_{1}(x)$ and $\rho_{2}(x)$
which can be imaged simultaneously~\cite{PhysRevA.80.023603}. The
bubble formation can be observed by plotting the normalized relative
number density distribution $p_{z}(x)=(\rho_{2}(x)-\rho_{1}(x))/(\rho_{2}(x)+\rho_{1}(x))$
versus evolution time.

\section{Conclusions\label{sec:Conclusions}}

We have considered a model that implements the quantum decay of a
relativistic false vacuum with an ultra-cold two-component spinor
Bose gas. The experimental realization appears to be feasible. If
achieved it will simulate quantum tunneling dynamics, potentially
in regimes that are not readily accessible either with analytic approximations
or numerical simulations on computers. This will provide a strong
test of our understanding of false vacuum decay in a cosmologically
relevant scenario. Beyond the features of Coleman's model, the ultra-cold
atom system has a tunable coupling between the relative and total
phase sectors of excitations. This has interesting consequences for
the scaling laws of the tunneling rate according to our prediction.

\textbf{Acknowledgments:} This work has been supported by the Marsden
Fund of New Zealand (contract Nos.~MAU1205 and UOO1320), the Australian
Research Council, the National Science Foundation under Grant No.~PHYS-1066293
and the hospitality of the Aspen Center for Physics. We thank R. Hulet
for scattering length data.

\bibliography{Vacuumquantumtunneling}

\begin{thebibliography}{49}%
\makeatletter
\providecommand \@ifxundefined [1]{%
 \@ifx{#1\undefined}
}%
\providecommand \@ifnum [1]{%
 \ifnum #1\expandafter \@firstoftwo
 \else \expandafter \@secondoftwo
 \fi
}%
\providecommand \@ifx [1]{%
 \ifx #1\expandafter \@firstoftwo
 \else \expandafter \@secondoftwo
 \fi
}%
\providecommand \natexlab [1]{#1}%
\providecommand \enquote  [1]{``#1''}%
\providecommand \bibnamefont  [1]{#1}%
\providecommand \bibfnamefont [1]{#1}%
\providecommand \citenamefont [1]{#1}%
\providecommand \href@noop [0]{\@secondoftwo}%
\providecommand \href [0]{\begingroup \@sanitize@url \@href}%
\providecommand \@href[1]{\@@startlink{#1}\@@href}%
\providecommand \@@href[1]{\endgroup#1\@@endlink}%
\providecommand \@sanitize@url [0]{\catcode `\\12\catcode `\$12\catcode
  `\&12\catcode `\#12\catcode `\^12\catcode `\_12\catcode `\%12\relax}%
\providecommand \@@startlink[1]{}%
\providecommand \@@endlink[0]{}%
\providecommand \url  [0]{\begingroup\@sanitize@url \@url }%
\providecommand \@url [1]{\endgroup\@href {#1}{\urlprefix }}%
\providecommand \urlprefix  [0]{URL }%
\providecommand \Eprint [0]{\href }%
\providecommand \doibase [0]{http://dx.doi.org/}%
\providecommand \selectlanguage [0]{\@gobble}%
\providecommand \bibinfo  [0]{\@secondoftwo}%
\providecommand \bibfield  [0]{\@secondoftwo}%
\providecommand \translation [1]{[#1]}%
\providecommand \BibitemOpen [0]{}%
\providecommand \bibitemStop [0]{}%
\providecommand \bibitemNoStop [0]{.\EOS\space}%
\providecommand \EOS [0]{\spacefactor3000\relax}%
\providecommand \BibitemShut  [1]{\csname bibitem#1\endcsname}%
\let\auto@bib@innerbib\@empty
\bibitem [{\citenamefont {Schmelzer}(2005)}]{Schmelzer2005}%
  \BibitemOpen
  \bibinfo {editor} {\bibfnamefont {J.~W.~P.}\ \bibnamefont {Schmelzer}},\
  ed.,\ \href {\doibase 10.1002/3527604790} {\emph {\bibinfo {title}
  {{Nucleation Theory and Applications}}}}\ (\bibinfo  {publisher} {Wiley-VCH
  Verlag GmbH \& Co. KGaA},\ \bibinfo {address} {Weinheim, FRG},\ \bibinfo
  {year} {2005})\BibitemShut {NoStop}%
\bibitem [{\citenamefont {Rayleigh}(1917)}]{Rayleigh17}%
  \BibitemOpen
  \bibfield  {author} {\bibinfo {author} {\bibfnamefont {L.}~\bibnamefont
  {Rayleigh}},\ }\href@noop {} {\bibfield  {journal} {\bibinfo  {journal}
  {Phil. Mag.}\ }\textbf {\bibinfo {volume} {34}},\ \bibinfo {pages} {94}
  (\bibinfo {year} {1917})}\BibitemShut {NoStop}%
\bibitem [{\citenamefont {Lifshitz}\ and\ \citenamefont
  {Kagan}(1972)}]{Lifshitz72}%
  \BibitemOpen
  \bibfield  {author} {\bibinfo {author} {\bibfnamefont {I.~M.}\ \bibnamefont
  {Lifshitz}}\ and\ \bibinfo {author} {\bibfnamefont {Y.}~\bibnamefont
  {Kagan}},\ }\href@noop {} {\bibfield  {journal} {\bibinfo  {journal} {Sov.
  Phys. JETP}\ }\textbf {\bibinfo {volume} {35}},\ \bibinfo {pages} {206}
  (\bibinfo {year} {1972})}\BibitemShut {NoStop}%
\bibitem [{\citenamefont {Satoh}\ \emph {et~al.}(1992)\citenamefont {Satoh},
  \citenamefont {Morishita}, \citenamefont {Ogata},\ and\ \citenamefont
  {Katoh}}]{Satoh92}%
  \BibitemOpen
  \bibfield  {author} {\bibinfo {author} {\bibfnamefont {T.}~\bibnamefont
  {Satoh}}, \bibinfo {author} {\bibfnamefont {M.}~\bibnamefont {Morishita}},
  \bibinfo {author} {\bibfnamefont {M.}~\bibnamefont {Ogata}}, \ and\ \bibinfo
  {author} {\bibfnamefont {S.}~\bibnamefont {Katoh}},\ }\href@noop {}
  {\bibfield  {journal} {\bibinfo  {journal} {Phys. Rev. Lett.}\ }\textbf
  {\bibinfo {volume} {69}},\ \bibinfo {pages} {335} (\bibinfo {year}
  {1992})}\BibitemShut {NoStop}%
\bibitem [{\citenamefont {Coleman}(1977)}]{Coleman1977}%
  \BibitemOpen
  \bibfield  {author} {\bibinfo {author} {\bibfnamefont {S.}~\bibnamefont
  {Coleman}},\ }\href {\doibase 10.1103/PhysRevD.15.2929} {\bibfield  {journal}
  {\bibinfo  {journal} {Phys. Rev. D}\ }\textbf {\bibinfo {volume} {15}},\
  \bibinfo {pages} {2929} (\bibinfo {year} {1977})}\BibitemShut {NoStop}%
\bibitem [{\citenamefont {Vilenkin}(1983)}]{Vilenkin1983}%
  \BibitemOpen
  \bibfield  {author} {\bibinfo {author} {\bibfnamefont {A.}~\bibnamefont
  {Vilenkin}},\ }\href {\doibase 10.1103/PhysRevD.27.2848} {\bibfield
  {journal} {\bibinfo  {journal} {Phys. Rev. D}\ }\textbf {\bibinfo {volume}
  {27}},\ \bibinfo {pages} {2848} (\bibinfo {year} {1983})}\BibitemShut
  {NoStop}%
\bibitem [{\citenamefont {Guth}(2007)}]{Guth2007}%
  \BibitemOpen
  \bibfield  {author} {\bibinfo {author} {\bibfnamefont {A.~H.}\ \bibnamefont
  {Guth}},\ }\href {\doibase 10.1088/1751-8113/40/25/S25} {\bibfield  {journal}
  {\bibinfo  {journal} {J. Phys. A: Math. Theor.}\ }\textbf {\bibinfo {volume}
  {40}},\ \bibinfo {pages} {6811} (\bibinfo {year} {2007})}\BibitemShut
  {NoStop}%
\bibitem [{\citenamefont {Coleman}\ and\ \citenamefont {{De
  Luccia}}(1980)}]{Coleman1980}%
  \BibitemOpen
  \bibfield  {author} {\bibinfo {author} {\bibfnamefont {S.}~\bibnamefont
  {Coleman}}\ and\ \bibinfo {author} {\bibfnamefont {F.}~\bibnamefont {{De
  Luccia}}},\ }\href {\doibase 10.1103/PhysRevD.21.3305} {\bibfield  {journal}
  {\bibinfo  {journal} {Phys. Rev. D}\ }\textbf {\bibinfo {volume} {21}},\
  \bibinfo {pages} {3305} (\bibinfo {year} {1980})}\BibitemShut {NoStop}%
\bibitem [{\citenamefont {Feeney}\ \emph {et~al.}(2011)\citenamefont {Feeney},
  \citenamefont {Johnson}, \citenamefont {Mortlock},\ and\ \citenamefont
  {Peiris}}]{Feeney2011-inflation}%
  \BibitemOpen
  \bibfield  {author} {\bibinfo {author} {\bibfnamefont {S.~M.}\ \bibnamefont
  {Feeney}}, \bibinfo {author} {\bibfnamefont {M.~C.}\ \bibnamefont {Johnson}},
  \bibinfo {author} {\bibfnamefont {D.~J.}\ \bibnamefont {Mortlock}}, \ and\
  \bibinfo {author} {\bibfnamefont {H.~V.}\ \bibnamefont {Peiris}},\ }\href
  {\doibase 10.1103/PhysRevLett.107.071301} {\bibfield  {journal} {\bibinfo
  {journal} {Phys. Rev. Lett.}\ }\textbf {\bibinfo {volume} {107}},\ \bibinfo
  {pages} {071301} (\bibinfo {year} {2011})}\BibitemShut {NoStop}%
\bibitem [{\citenamefont {Bousso}\ \emph {et~al.}(2013)\citenamefont {Bousso},
  \citenamefont {Harlow},\ and\ \citenamefont {Senatore}}]{Bousso2013}%
  \BibitemOpen
  \bibfield  {author} {\bibinfo {author} {\bibfnamefont {R.}~\bibnamefont
  {Bousso}}, \bibinfo {author} {\bibfnamefont {D.}~\bibnamefont {Harlow}}, \
  and\ \bibinfo {author} {\bibfnamefont {L.}~\bibnamefont {Senatore}},\ }\href
  {http://arxiv.org/abs/1309.4060} {\enquote {\bibinfo {title} {{Inflation
  after False Vacuum Decay: Observational Prospects after Planck}},}\ }\bibinfo
  {howpublished} {arXiv:1309.4060} (\bibinfo {year} {2013})\BibitemShut
  {NoStop}%
\bibitem [{\citenamefont {Callan}\ and\ \citenamefont
  {Coleman}(1977)}]{Callan1977}%
  \BibitemOpen
  \bibfield  {author} {\bibinfo {author} {\bibfnamefont {C.~G.}\ \bibnamefont
  {Callan}}\ and\ \bibinfo {author} {\bibfnamefont {S.}~\bibnamefont
  {Coleman}},\ }\href {\doibase 10.1103/PhysRevD.16.1762} {\bibfield  {journal}
  {\bibinfo  {journal} {Phys. Rev. D}\ }\textbf {\bibinfo {volume} {16}},\
  \bibinfo {pages} {1762} (\bibinfo {year} {1977})}\BibitemShut {NoStop}%
\bibitem [{\citenamefont {Fialko}\ \emph {et~al.}(2015)\citenamefont {Fialko},
  \citenamefont {Opanchuk}, \citenamefont {Sidorov}, \citenamefont {Drummond},\
  and\ \citenamefont {Brand}}]{Fialko15}%
  \BibitemOpen
  \bibfield  {author} {\bibinfo {author} {\bibfnamefont {O.}~\bibnamefont
  {Fialko}}, \bibinfo {author} {\bibfnamefont {B.}~\bibnamefont {Opanchuk}},
  \bibinfo {author} {\bibfnamefont {A.}~\bibnamefont {Sidorov}}, \bibinfo
  {author} {\bibfnamefont {P.}~\bibnamefont {Drummond}}, \ and\ \bibinfo
  {author} {\bibfnamefont {J.}~\bibnamefont {Brand}},\ }\href@noop {}
  {\bibfield  {journal} {\bibinfo  {journal} {Eur. Phys. Lett.}\ }\textbf
  {\bibinfo {volume} {110}},\ \bibinfo {pages} {56001} (\bibinfo {year}
  {2015})}\BibitemShut {NoStop}%
\bibitem [{\citenamefont {Egorov}\ \emph {et~al.}(2011)\citenamefont {Egorov}
  \emph {et~al.}}]{Egorov2011}%
  \BibitemOpen
  \bibfield  {author} {\bibinfo {author} {\bibfnamefont {M.}~\bibnamefont
  {Egorov}} \emph {et~al.},\ }\href {\doibase 10.1103/PhysRevA.84.021605}
  {\bibfield  {journal} {\bibinfo  {journal} {Phys. Rev. A}\ }\textbf {\bibinfo
  {volume} {84}},\ \bibinfo {pages} {021605(R)} (\bibinfo {year}
  {2011})}\BibitemShut {NoStop}%
\bibitem [{\citenamefont {Opanchuk}\ \emph {et~al.}(2013)\citenamefont
  {Opanchuk}, \citenamefont {Polkinghorne}, \citenamefont {Fialko},
  \citenamefont {Brand},\ and\ \citenamefont
  {Drummond}}]{Opanchuk2013-early-universe}%
  \BibitemOpen
  \bibfield  {author} {\bibinfo {author} {\bibfnamefont {B.}~\bibnamefont
  {Opanchuk}}, \bibinfo {author} {\bibfnamefont {R.}~\bibnamefont
  {Polkinghorne}}, \bibinfo {author} {\bibfnamefont {O.}~\bibnamefont
  {Fialko}}, \bibinfo {author} {\bibfnamefont {J.}~\bibnamefont {Brand}}, \
  and\ \bibinfo {author} {\bibfnamefont {P.~D.}\ \bibnamefont {Drummond}},\
  }\href {\doibase 10.1002/andp.201300113} {\bibfield  {journal} {\bibinfo
  {journal} {Ann. Phys.}\ }\textbf {\bibinfo {volume} {525}},\ \bibinfo {pages}
  {866} (\bibinfo {year} {2013})}\BibitemShut {NoStop}%
\bibitem [{\citenamefont {Kapitza}(1951)}]{Kapitza1951-JETP}%
  \BibitemOpen
  \bibfield  {author} {\bibinfo {author} {\bibfnamefont {P.~L.}\ \bibnamefont
  {Kapitza}},\ }\href@noop {} {\bibfield  {journal} {\bibinfo  {journal} {Sov.
  Phys. JETP}\ }\textbf {\bibinfo {volume} {21}},\ \bibinfo {pages} {588}
  (\bibinfo {year} {1951})},\ \bibinfo {note} {for previous applications of the
  idea to BECs see H.\ Saito and M.\ Ueda, Phys.\ Rev.\ Lett.\ \textbf{90},
  040403 (2003) and H.\ Saito, R.\ G.\ Hulet, and M.\ Ueda, Phys.\ Rev.\ A
  \textbf{76}, 053619 (2007)}\BibitemShut {NoStop}%
\bibitem [{\citenamefont {Citro}\ \emph {et~al.}(2015)\citenamefont {Citro},
  \citenamefont {{Dalla Torre}}, \citenamefont {D'Alessio}, \citenamefont
  {Polkovnikov}, \citenamefont {Babadi}, \citenamefont {Oka},\ and\
  \citenamefont {Demler}}]{Citro2015a}%
  \BibitemOpen
  \bibfield  {author} {\bibinfo {author} {\bibfnamefont {R.}~\bibnamefont
  {Citro}}, \bibinfo {author} {\bibfnamefont {E.~G.}\ \bibnamefont {{Dalla
  Torre}}}, \bibinfo {author} {\bibfnamefont {L.}~\bibnamefont {D'Alessio}},
  \bibinfo {author} {\bibfnamefont {A.}~\bibnamefont {Polkovnikov}}, \bibinfo
  {author} {\bibfnamefont {M.}~\bibnamefont {Babadi}}, \bibinfo {author}
  {\bibfnamefont {T.}~\bibnamefont {Oka}}, \ and\ \bibinfo {author}
  {\bibfnamefont {E.}~\bibnamefont {Demler}},\ }\href {\doibase
  10.1016/j.aop.2015.03.027} {\bibfield  {journal} {\bibinfo  {journal} {Ann.
  Phys. (N. Y).}\ }\textbf {\bibinfo {volume} {360}},\ \bibinfo {pages} {694}
  (\bibinfo {year} {2015})}\BibitemShut {NoStop}%
\bibitem [{\citenamefont {Caldeira}\ and\ \citenamefont
  {Leggett}(1981)}]{Caldeira1981}%
  \BibitemOpen
  \bibfield  {author} {\bibinfo {author} {\bibfnamefont {A.~O.}\ \bibnamefont
  {Caldeira}}\ and\ \bibinfo {author} {\bibfnamefont {A.~J.}\ \bibnamefont
  {Leggett}},\ }\href {\doibase 10.1103/PhysRevLett.46.211} {\bibfield
  {journal} {\bibinfo  {journal} {Phys. Rev. Lett.}\ }\textbf {\bibinfo
  {volume} {46}},\ \bibinfo {pages} {211} (\bibinfo {year} {1981})}\BibitemShut
  {NoStop}%
\bibitem [{\citenamefont {Fischer}\ and\ \citenamefont
  {Sch\"{u}tzhold}(2004)}]{Fischer2004}%
  \BibitemOpen
  \bibfield  {author} {\bibinfo {author} {\bibfnamefont {U.~R.}\ \bibnamefont
  {Fischer}}\ and\ \bibinfo {author} {\bibfnamefont {R.}~\bibnamefont
  {Sch\"{u}tzhold}},\ }\href {\doibase 10.1103/PhysRevA.70.063615} {\bibfield
  {journal} {\bibinfo  {journal} {Phys. Rev. A}\ }\textbf {\bibinfo {volume}
  {70}},\ \bibinfo {pages} {063615} (\bibinfo {year} {2004})}\BibitemShut
  {NoStop}%
\bibitem [{\citenamefont {Menicucci}\ \emph {et~al.}(2010)\citenamefont
  {Menicucci}, \citenamefont {Olson},\ and\ \citenamefont
  {Milburn}}]{Menicucci2010}%
  \BibitemOpen
  \bibfield  {author} {\bibinfo {author} {\bibfnamefont {N.~C.}\ \bibnamefont
  {Menicucci}}, \bibinfo {author} {\bibfnamefont {S.~J.}\ \bibnamefont
  {Olson}}, \ and\ \bibinfo {author} {\bibfnamefont {G.~J.}\ \bibnamefont
  {Milburn}},\ }\href {\doibase 10.1088/1367-2630/12/9/095019} {\bibfield
  {journal} {\bibinfo  {journal} {New J. Phys.}\ }\textbf {\bibinfo {volume}
  {12}},\ \bibinfo {pages} {095019} (\bibinfo {year} {2010})}\BibitemShut
  {NoStop}%
\bibitem [{\citenamefont {Neuenhahn}\ \emph {et~al.}(2012)\citenamefont
  {Neuenhahn}, \citenamefont {Polkovnikov},\ and\ \citenamefont
  {Marquardt}}]{Neuenhahn2012-phase-structures}%
  \BibitemOpen
  \bibfield  {author} {\bibinfo {author} {\bibfnamefont {C.}~\bibnamefont
  {Neuenhahn}}, \bibinfo {author} {\bibfnamefont {A.}~\bibnamefont
  {Polkovnikov}}, \ and\ \bibinfo {author} {\bibfnamefont {F.}~\bibnamefont
  {Marquardt}},\ }\href {\doibase 10.1103/PhysRevLett.109.085304} {\bibfield
  {journal} {\bibinfo  {journal} {Phys. Rev. Lett.}\ }\textbf {\bibinfo
  {volume} {109}},\ \bibinfo {pages} {085304} (\bibinfo {year}
  {2012})}\BibitemShut {NoStop}%
\bibitem [{\citenamefont {Su}\ \emph {et~al.}(2015)\citenamefont {Su},
  \citenamefont {Gou}, \citenamefont {Liu}, \citenamefont {Bradley},
  \citenamefont {Fialko},\ and\ \citenamefont {Brand}}]{Su2014a}%
  \BibitemOpen
  \bibfield  {author} {\bibinfo {author} {\bibfnamefont {S.-W.}\ \bibnamefont
  {Su}}, \bibinfo {author} {\bibfnamefont {S.-C.}\ \bibnamefont {Gou}},
  \bibinfo {author} {\bibfnamefont {I.-K.}\ \bibnamefont {Liu}}, \bibinfo
  {author} {\bibfnamefont {A.~S.}\ \bibnamefont {Bradley}}, \bibinfo {author}
  {\bibfnamefont {O.}~\bibnamefont {Fialko}}, \ and\ \bibinfo {author}
  {\bibfnamefont {J.}~\bibnamefont {Brand}},\ }\href {\doibase
  10.1103/PhysRevA.91.023631} {\bibfield  {journal} {\bibinfo  {journal} {Phys.
  Rev. A}\ }\textbf {\bibinfo {volume} {91}},\ \bibinfo {pages} {023631}
  (\bibinfo {year} {2015})}\BibitemShut {NoStop}%
\bibitem [{\citenamefont {Liddle}\ and\ \citenamefont
  {Lyth}(2000)}]{Liddle2000}%
  \BibitemOpen
  \bibfield  {author} {\bibinfo {author} {\bibfnamefont {A.~R.}\ \bibnamefont
  {Liddle}}\ and\ \bibinfo {author} {\bibfnamefont {D.~H.}\ \bibnamefont
  {Lyth}},\ }\href
  {http://www.cambridge.org/au/academic/subjects/physics/cosmology-relativity-and-gravitation/cosmological-inflation-and-large-scale-structure?format=PB}
  {\emph {\bibinfo {title} {{Cosmological Inflation and Large-Scale
  Structure}}}}\ (\bibinfo  {publisher} {Cambridge University Press},\ \bibinfo
  {year} {2000})\ p.\ \bibinfo {pages} {400}\BibitemShut {NoStop}%
\bibitem [{\citenamefont {Amin}\ \emph {et~al.}(2012)\citenamefont {Amin},
  \citenamefont {Easther}, \citenamefont {Finkel}, \citenamefont {Flauger},\
  and\ \citenamefont {Hertzberg}}]{Amin2012}%
  \BibitemOpen
  \bibfield  {author} {\bibinfo {author} {\bibfnamefont {M.~A.}\ \bibnamefont
  {Amin}}, \bibinfo {author} {\bibfnamefont {R.}~\bibnamefont {Easther}},
  \bibinfo {author} {\bibfnamefont {H.}~\bibnamefont {Finkel}}, \bibinfo
  {author} {\bibfnamefont {R.}~\bibnamefont {Flauger}}, \ and\ \bibinfo
  {author} {\bibfnamefont {M.~P.}\ \bibnamefont {Hertzberg}},\ }\href {\doibase
  10.1103/PhysRevLett.108.241302} {\bibfield  {journal} {\bibinfo  {journal}
  {Phys. Rev. Lett.}\ }\textbf {\bibinfo {volume} {108}},\ \bibinfo {pages}
  {241302} (\bibinfo {year} {2012})}\BibitemShut {NoStop}%
\bibitem [{\citenamefont {Drummond}\ and\ \citenamefont
  {Hardman}(1993)}]{Drummond1993a}%
  \BibitemOpen
  \bibfield  {author} {\bibinfo {author} {\bibfnamefont {P.~D.}\ \bibnamefont
  {Drummond}}\ and\ \bibinfo {author} {\bibfnamefont {A.~D.}\ \bibnamefont
  {Hardman}},\ }\href@noop {} {\bibfield  {journal} {\bibinfo  {journal}
  {EuroPhys. Lett.}\ }\textbf {\bibinfo {volume} {21}},\ \bibinfo {pages} {279}
  (\bibinfo {year} {1993})}\BibitemShut {NoStop}%
\bibitem [{\citenamefont {Steel}\ \emph {et~al.}(1998)\citenamefont {Steel}
  \emph {et~al.}}]{Steel1998}%
  \BibitemOpen
  \bibfield  {author} {\bibinfo {author} {\bibfnamefont {M.}~\bibnamefont
  {Steel}} \emph {et~al.},\ }\href {\doibase 10.1103/PhysRevA.58.4824}
  {\bibfield  {journal} {\bibinfo  {journal} {Phys. Rev. A}\ }\textbf {\bibinfo
  {volume} {58}},\ \bibinfo {pages} {4824} (\bibinfo {year}
  {1998})}\BibitemShut {NoStop}%
\bibitem [{\citenamefont {Sinatra}\ \emph {et~al.}(2002)\citenamefont
  {Sinatra}, \citenamefont {Lobo},\ and\ \citenamefont {Castin}}]{Sinatra2002}%
  \BibitemOpen
  \bibfield  {author} {\bibinfo {author} {\bibfnamefont {A.}~\bibnamefont
  {Sinatra}}, \bibinfo {author} {\bibfnamefont {C.}~\bibnamefont {Lobo}}, \
  and\ \bibinfo {author} {\bibfnamefont {Y.}~\bibnamefont {Castin}},\ }\href
  {\doibase 10.1088/0953-4075/35/17/301} {\bibfield  {journal} {\bibinfo
  {journal} {J. Phys. B}\ }\textbf {\bibinfo {volume} {35}},\ \bibinfo {pages}
  {3599} (\bibinfo {year} {2002})}\BibitemShut {NoStop}%
\bibitem [{\citenamefont {Drummond}\ and\ \citenamefont
  {Chaturvedi}(2016)}]{Drummond2016}%
  \BibitemOpen
  \bibfield  {author} {\bibinfo {author} {\bibfnamefont {P.~D.}\ \bibnamefont
  {Drummond}}\ and\ \bibinfo {author} {\bibfnamefont {S.}~\bibnamefont
  {Chaturvedi}},\ }\href@noop {} {\bibfield  {journal} {\bibinfo  {journal}
  {Physica Scripta}\ }\textbf {\bibinfo {volume} {91}},\ \bibinfo {pages}
  {073007} (\bibinfo {year} {2016})}\BibitemShut {NoStop}%
\bibitem [{\citenamefont {Drummond}\ and\ \citenamefont
  {Kinsler}(1989)}]{Drummond1989}%
  \BibitemOpen
  \bibfield  {author} {\bibinfo {author} {\bibfnamefont {P.~D.}\ \bibnamefont
  {Drummond}}\ and\ \bibinfo {author} {\bibfnamefont {P.}~\bibnamefont
  {Kinsler}},\ }\href {\doibase 10.1103/PhysRevA.40.4813} {\bibfield  {journal}
  {\bibinfo  {journal} {Phys. Rev. A}\ }\textbf {\bibinfo {volume} {40}},\
  \bibinfo {pages} {4813} (\bibinfo {year} {1989})}\BibitemShut {NoStop}%
\bibitem [{\citenamefont {Anderson}\ \emph {et~al.}(2009)\citenamefont
  {Anderson}, \citenamefont {Ticknor}, \citenamefont {Sidorov},\ and\
  \citenamefont {Hall}}]{PhysRevA.80.023603}%
  \BibitemOpen
  \bibfield  {author} {\bibinfo {author} {\bibfnamefont {R.~P.}\ \bibnamefont
  {Anderson}}, \bibinfo {author} {\bibfnamefont {C.}~\bibnamefont {Ticknor}},
  \bibinfo {author} {\bibfnamefont {A.~I.}\ \bibnamefont {Sidorov}}, \ and\
  \bibinfo {author} {\bibfnamefont {B.~V.}\ \bibnamefont {Hall}},\ }\href
  {\doibase 10.1103/PhysRevA.80.023603} {\bibfield  {journal} {\bibinfo
  {journal} {Phys. Rev. A}\ }\textbf {\bibinfo {volume} {80}},\ \bibinfo
  {pages} {023603} (\bibinfo {year} {2009})}\BibitemShut {NoStop}%
\bibitem [{\citenamefont {Goldman}\ and\ \citenamefont
  {Dalibard}(2014)}]{goldman14}%
  \BibitemOpen
  \bibfield  {author} {\bibinfo {author} {\bibfnamefont {N.}~\bibnamefont
  {Goldman}}\ and\ \bibinfo {author} {\bibfnamefont {J.}~\bibnamefont
  {Dalibard}},\ }\href@noop {} {\bibfield  {journal} {\bibinfo  {journal}
  {Physical Review X}\ }\textbf {\bibinfo {volume} {4}},\ \bibinfo {pages}
  {031027} (\bibinfo {year} {2014})}\BibitemShut {NoStop}%
\bibitem [{\citenamefont {Atland}\ and\ \citenamefont
  {Simons}(2010)}]{Atland2010}%
  \BibitemOpen
  \bibfield  {author} {\bibinfo {author} {\bibfnamefont {A.}~\bibnamefont
  {Atland}}\ and\ \bibinfo {author} {\bibfnamefont {B.}~\bibnamefont
  {Simons}},\ }\href@noop {} {\emph {\bibinfo {title} {{Condensed Matter Field
  Theory}}}}\ (\bibinfo  {publisher} {Cambridge University Press},\ \bibinfo
  {year} {2010})\ p.\ \bibinfo {pages} {783}\BibitemShut {NoStop}%
\bibitem [{\citenamefont {Takagi}(2006)}]{Takagi2006}%
  \BibitemOpen
  \bibfield  {author} {\bibinfo {author} {\bibfnamefont {S.}~\bibnamefont
  {Takagi}},\ }\href@noop {} {\emph {\bibinfo {title} {{Macroscopic Quantum
  Tunneling}}}}\ (\bibinfo  {publisher} {Cambridge University Press},\ \bibinfo
  {year} {2006})\ p.\ \bibinfo {pages} {224}\BibitemShut {NoStop}%
\bibitem [{\citenamefont {Danshita}\ and\ \citenamefont
  {Polkovnikov}(2012)}]{Danshita12}%
  \BibitemOpen
  \bibfield  {author} {\bibinfo {author} {\bibfnamefont {I.}~\bibnamefont
  {Danshita}}\ and\ \bibinfo {author} {\bibfnamefont {A.}~\bibnamefont
  {Polkovnikov}},\ }\href@noop {} {\bibfield  {journal} {\bibinfo  {journal}
  {Phys. Rev. A}\ }\textbf {\bibinfo {volume} {85}},\ \bibinfo {pages} {023638}
  (\bibinfo {year} {2012})}\BibitemShut {NoStop}%
\bibitem [{\citenamefont {Deuar}\ and\ \citenamefont
  {Drummond}(2006)}]{Deuar:2006}%
  \BibitemOpen
  \bibfield  {author} {\bibinfo {author} {\bibfnamefont {P.}~\bibnamefont
  {Deuar}}\ and\ \bibinfo {author} {\bibfnamefont {P.~D.}\ \bibnamefont
  {Drummond}},\ }\href@noop {} {\bibfield  {journal} {\bibinfo  {journal}
  {Journal of Physics A}\ }\textbf {\bibinfo {volume} {39}},\ \bibinfo {pages}
  {2723 } (\bibinfo {year} {2006})}\BibitemShut {NoStop}%
\bibitem [{\citenamefont {Carter}\ \emph {et~al.}(1987)\citenamefont {Carter},
  \citenamefont {Drummond}, \citenamefont {Reid},\ and\ \citenamefont
  {Shelby}}]{Carter1987a}%
  \BibitemOpen
  \bibfield  {author} {\bibinfo {author} {\bibfnamefont {S.~J.}\ \bibnamefont
  {Carter}}, \bibinfo {author} {\bibfnamefont {P.~D.}\ \bibnamefont
  {Drummond}}, \bibinfo {author} {\bibfnamefont {M.~D.}\ \bibnamefont {Reid}},
  \ and\ \bibinfo {author} {\bibfnamefont {R.~M.}\ \bibnamefont {Shelby}},\
  }\href@noop {} {\bibfield  {journal} {\bibinfo  {journal} {Phys. Rev. Lett.}\
  }\textbf {\bibinfo {volume} {58}},\ \bibinfo {pages} {1841} (\bibinfo {year}
  {1987})}\BibitemShut {NoStop}%
\bibitem [{\citenamefont {Drummond}\ \emph {et~al.}(1993)\citenamefont
  {Drummond}, \citenamefont {Shelby}, \citenamefont {Friberg},\ and\
  \citenamefont {Yamamoto}}]{Drummond1993b}%
  \BibitemOpen
  \bibfield  {author} {\bibinfo {author} {\bibfnamefont {P.~D.}\ \bibnamefont
  {Drummond}}, \bibinfo {author} {\bibfnamefont {R.~M.}\ \bibnamefont
  {Shelby}}, \bibinfo {author} {\bibfnamefont {S.~R.}\ \bibnamefont {Friberg}},
  \ and\ \bibinfo {author} {\bibfnamefont {Y.}~\bibnamefont {Yamamoto}},\
  }\href@noop {} {\bibfield  {journal} {\bibinfo  {journal} {Nature}\ }\textbf
  {\bibinfo {volume} {365}},\ \bibinfo {pages} {307} (\bibinfo {year}
  {1993})}\BibitemShut {NoStop}%
\bibitem [{\citenamefont {Corney}\ \emph {et~al.}(2006)\citenamefont {Corney},
  \citenamefont {Drummond}, \citenamefont {Heersink}, \citenamefont {Josse},
  \citenamefont {Leuchs},\ and\ \citenamefont {Andersen}}]{Corney2006c}%
  \BibitemOpen
  \bibfield  {author} {\bibinfo {author} {\bibfnamefont {J.~F.}\ \bibnamefont
  {Corney}}, \bibinfo {author} {\bibfnamefont {P.~D.}\ \bibnamefont
  {Drummond}}, \bibinfo {author} {\bibfnamefont {J.}~\bibnamefont {Heersink}},
  \bibinfo {author} {\bibfnamefont {V.}~\bibnamefont {Josse}}, \bibinfo
  {author} {\bibfnamefont {G.}~\bibnamefont {Leuchs}}, \ and\ \bibinfo {author}
  {\bibfnamefont {U.~L.}\ \bibnamefont {Andersen}},\ }\href@noop {} {\bibfield
  {journal} {\bibinfo  {journal} {Phys. Rev. Lett.}\ }\textbf {\bibinfo
  {volume} {97}},\ \bibinfo {pages} {023606} (\bibinfo {year}
  {2006})}\BibitemShut {NoStop}%
\bibitem [{\citenamefont {Deuar}\ and\ \citenamefont
  {Drummond}(2007)}]{Deuar:2007_BECCollisions}%
  \BibitemOpen
  \bibfield  {author} {\bibinfo {author} {\bibfnamefont {P.}~\bibnamefont
  {Deuar}}\ and\ \bibinfo {author} {\bibfnamefont {P.~D.}\ \bibnamefont
  {Drummond}},\ }\href@noop {} {\bibfield  {journal} {\bibinfo  {journal}
  {Phys. Rev. Lett.}\ }\textbf {\bibinfo {volume} {98}},\ \bibinfo {pages}
  {120402} (\bibinfo {year} {2007})}\BibitemShut {NoStop}%
\bibitem [{\citenamefont {Bagnato}\ and\ \citenamefont
  {Kleppner}(1991)}]{Bagnato91}%
  \BibitemOpen
  \bibfield  {author} {\bibinfo {author} {\bibfnamefont {V.}~\bibnamefont
  {Bagnato}}\ and\ \bibinfo {author} {\bibfnamefont {D.}~\bibnamefont
  {Kleppner}},\ }\href@noop {} {\bibfield  {journal} {\bibinfo  {journal}
  {Phys. Rev. A}\ }\textbf {\bibinfo {volume} {44}},\ \bibinfo {pages} {7439}
  (\bibinfo {year} {1991})}\BibitemShut {NoStop}%
\bibitem [{\citenamefont {G\"orlitz}\ \emph {et~al.}(2001)\citenamefont
  {G\"orlitz}, \citenamefont {Vogels}, \citenamefont {Leanhardt}, \citenamefont
  {Raman}, \citenamefont {Gustavson}, \citenamefont {Abo-Shaeer}, \citenamefont
  {Chikkatur}, \citenamefont {Gupta}, \citenamefont {Inouye}, \citenamefont
  {Rosenband},\ and\ \citenamefont {Ketterle}}]{Ketterle01}%
  \BibitemOpen
  \bibfield  {author} {\bibinfo {author} {\bibfnamefont {A.}~\bibnamefont
  {G\"orlitz}}, \bibinfo {author} {\bibfnamefont {J.~M.}\ \bibnamefont
  {Vogels}}, \bibinfo {author} {\bibfnamefont {A.~E.}\ \bibnamefont
  {Leanhardt}}, \bibinfo {author} {\bibfnamefont {C.}~\bibnamefont {Raman}},
  \bibinfo {author} {\bibfnamefont {T.~L.}\ \bibnamefont {Gustavson}}, \bibinfo
  {author} {\bibfnamefont {J.~R.}\ \bibnamefont {Abo-Shaeer}}, \bibinfo
  {author} {\bibfnamefont {A.~P.}\ \bibnamefont {Chikkatur}}, \bibinfo {author}
  {\bibfnamefont {S.}~\bibnamefont {Gupta}}, \bibinfo {author} {\bibfnamefont
  {S.}~\bibnamefont {Inouye}}, \bibinfo {author} {\bibfnamefont
  {T.}~\bibnamefont {Rosenband}}, \ and\ \bibinfo {author} {\bibfnamefont
  {W.}~\bibnamefont {Ketterle}},\ }\href@noop {} {\bibfield  {journal}
  {\bibinfo  {journal} {Phys. Rev. Lett.}\ }\textbf {\bibinfo {volume} {87}},\
  \bibinfo {pages} {130402} (\bibinfo {year} {2001})}\BibitemShut {NoStop}%
\bibitem [{\citenamefont {Hillery}\ \emph {et~al.}(1984)\citenamefont
  {Hillery}, \citenamefont {O'Connell}, \citenamefont {Scully},\ and\
  \citenamefont {Wigner}}]{Hillery_Review_1984_DistributionFunctions}%
  \BibitemOpen
  \bibfield  {author} {\bibinfo {author} {\bibfnamefont {M.}~\bibnamefont
  {Hillery}}, \bibinfo {author} {\bibfnamefont {R.~F.}\ \bibnamefont
  {O'Connell}}, \bibinfo {author} {\bibfnamefont {M.~O.}\ \bibnamefont
  {Scully}}, \ and\ \bibinfo {author} {\bibfnamefont {E.~P.}\ \bibnamefont
  {Wigner}},\ }\href@noop {} {\bibfield  {journal} {\bibinfo  {journal} {Phys.
  Rep.}\ }\textbf {\bibinfo {volume} {106}},\ \bibinfo {pages} {121} (\bibinfo
  {year} {1984})}\BibitemShut {NoStop}%
\bibitem [{\citenamefont {Lewis-Swan}\ \emph {et~al.}(2016)\citenamefont
  {Lewis-Swan}, \citenamefont {Olsen},\ and\ \citenamefont
  {Kheruntsyan}}]{lewis2016approximate}%
  \BibitemOpen
  \bibfield  {author} {\bibinfo {author} {\bibfnamefont {R.}~\bibnamefont
  {Lewis-Swan}}, \bibinfo {author} {\bibfnamefont {M.}~\bibnamefont {Olsen}}, \
  and\ \bibinfo {author} {\bibfnamefont {K.}~\bibnamefont {Kheruntsyan}},\
  }\href@noop {} {\bibfield  {journal} {\bibinfo  {journal} {arXiv preprint
  arXiv:1605.07276}\ } (\bibinfo {year} {2016})}\BibitemShut {NoStop}%
\bibitem [{\citenamefont {Ruostekoski}\ and\ \citenamefont
  {Isella}(2005)}]{Ruostekoski2005}%
  \BibitemOpen
  \bibfield  {author} {\bibinfo {author} {\bibfnamefont {J.}~\bibnamefont
  {Ruostekoski}}\ and\ \bibinfo {author} {\bibfnamefont {L.}~\bibnamefont
  {Isella}},\ }\href@noop {} {\bibfield  {journal} {\bibinfo  {journal} {Phys.
  Rev. Lett.}\ }\textbf {\bibinfo {volume} {95}},\ \bibinfo {pages} {110403}
  (\bibinfo {year} {2005})}\BibitemShut {NoStop}%
\bibitem [{\citenamefont {Lysebo}\ and\ \citenamefont
  {Veseth}(2010)}]{Lysebo2010}%
  \BibitemOpen
  \bibfield  {author} {\bibinfo {author} {\bibfnamefont {M.}~\bibnamefont
  {Lysebo}}\ and\ \bibinfo {author} {\bibfnamefont {L.}~\bibnamefont
  {Veseth}},\ }\href {\doibase 10.1103/PhysRevA.81.032702} {\bibfield
  {journal} {\bibinfo  {journal} {Phys. Rev. A}\ }\textbf {\bibinfo {volume}
  {81}},\ \bibinfo {pages} {032702} (\bibinfo {year} {2010})}\BibitemShut
  {NoStop}%
\bibitem [{\citenamefont {Hulet}()}]{Hulet2015}%
  \BibitemOpen
  \bibfield  {author} {\bibinfo {author} {\bibfnamefont {R.}~\bibnamefont
  {Hulet}},\ }\href@noop {} {}\bibinfo {note} {Private
  communication}\BibitemShut {NoStop}%
\bibitem [{\citenamefont {Bouchoule}\ \emph {et~al.}(2012)\citenamefont
  {Bouchoule}, \citenamefont {Arzamasovs}, \citenamefont {Kheruntsyan},\ and\
  \citenamefont {Gangardt}}]{Bouchoule2012-two-body}%
  \BibitemOpen
  \bibfield  {author} {\bibinfo {author} {\bibfnamefont {I.}~\bibnamefont
  {Bouchoule}}, \bibinfo {author} {\bibfnamefont {M.}~\bibnamefont
  {Arzamasovs}}, \bibinfo {author} {\bibfnamefont {K.~V.}\ \bibnamefont
  {Kheruntsyan}}, \ and\ \bibinfo {author} {\bibfnamefont {D.~M.}\ \bibnamefont
  {Gangardt}},\ }\href {\doibase 10.1103/PhysRevA.86.033626} {\bibfield
  {journal} {\bibinfo  {journal} {Phys. Rev. A}\ }\textbf {\bibinfo {volume}
  {86}},\ \bibinfo {pages} {033626} (\bibinfo {year} {2012})}\BibitemShut
  {NoStop}%
\bibitem [{\citenamefont {Goldman}\ \emph {et~al.}(2014)\citenamefont
  {Goldman}, \citenamefont {Juzeliunas}, \citenamefont {Ohberg},\ and\
  \citenamefont {Spielman}}]{spielman14}%
  \BibitemOpen
  \bibfield  {author} {\bibinfo {author} {\bibfnamefont {N.}~\bibnamefont
  {Goldman}}, \bibinfo {author} {\bibfnamefont {G.}~\bibnamefont {Juzeliunas}},
  \bibinfo {author} {\bibfnamefont {P.}~\bibnamefont {Ohberg}}, \ and\ \bibinfo
  {author} {\bibfnamefont {I.}~\bibnamefont {Spielman}},\ }\href@noop {}
  {\bibfield  {journal} {\bibinfo  {journal} {Rep. Prog. Phys}\ }\textbf
  {\bibinfo {volume} {77}},\ \bibinfo {pages} {126401} (\bibinfo {year}
  {2014})}\BibitemShut {NoStop}%
\bibitem [{\citenamefont {Jimenez-Garcia}\ \emph {et~al.}(2015)\citenamefont
  {Jimenez-Garcia}, \citenamefont {LeBlanc}, \citenamefont {Williams},
  \citenamefont {Beeler}, \citenamefont {Qu}, \citenamefont {Gong},
  \citenamefont {Zhang},\ and\ \citenamefont {Spielman}}]{spielman15}%
  \BibitemOpen
  \bibfield  {author} {\bibinfo {author} {\bibfnamefont {K.}~\bibnamefont
  {Jimenez-Garcia}}, \bibinfo {author} {\bibfnamefont {L.}~\bibnamefont
  {LeBlanc}}, \bibinfo {author} {\bibfnamefont {R.}~\bibnamefont {Williams}},
  \bibinfo {author} {\bibfnamefont {M.}~\bibnamefont {Beeler}}, \bibinfo
  {author} {\bibfnamefont {C.}~\bibnamefont {Qu}}, \bibinfo {author}
  {\bibfnamefont {M.}~\bibnamefont {Gong}}, \bibinfo {author} {\bibfnamefont
  {C.}~\bibnamefont {Zhang}}, \ and\ \bibinfo {author} {\bibfnamefont
  {I.}~\bibnamefont {Spielman}},\ }\href@noop {} {\bibfield  {journal}
  {\bibinfo  {journal} {Phys. Rev. Lett.}\ }\textbf {\bibinfo {volume} {114}},\
  \bibinfo {pages} {125301} (\bibinfo {year} {2015})}\BibitemShut {NoStop}%
\bibitem [{\citenamefont {Ramanathan}\ \emph {et~al.}(2011)\citenamefont
  {Ramanathan}, \citenamefont {Wright}, \citenamefont {Muniz}, \citenamefont
  {Zelan}, \citenamefont {Hill}, \citenamefont {Lobb}, \citenamefont
  {Helmerson}, \citenamefont {Phillips},\ and\ \citenamefont
  {Campbell}}]{PhysRevLett.106.130401}%
  \BibitemOpen
  \bibfield  {author} {\bibinfo {author} {\bibfnamefont {A.}~\bibnamefont
  {Ramanathan}}, \bibinfo {author} {\bibfnamefont {K.~C.}\ \bibnamefont
  {Wright}}, \bibinfo {author} {\bibfnamefont {S.~R.}\ \bibnamefont {Muniz}},
  \bibinfo {author} {\bibfnamefont {M.}~\bibnamefont {Zelan}}, \bibinfo
  {author} {\bibfnamefont {W.~T.}\ \bibnamefont {Hill}}, \bibinfo {author}
  {\bibfnamefont {C.~J.}\ \bibnamefont {Lobb}}, \bibinfo {author}
  {\bibfnamefont {K.}~\bibnamefont {Helmerson}}, \bibinfo {author}
  {\bibfnamefont {W.~D.}\ \bibnamefont {Phillips}}, \ and\ \bibinfo {author}
  {\bibfnamefont {G.~K.}\ \bibnamefont {Campbell}},\ }\href {\doibase
  10.1103/PhysRevLett.106.130401} {\bibfield  {journal} {\bibinfo  {journal}
  {Phys. Rev. Lett.}\ }\textbf {\bibinfo {volume} {106}},\ \bibinfo {pages}
  {130401} (\bibinfo {year} {2011})}\BibitemShut {NoStop}%
\end{thebibliography}%

\clearpage{}
\end{document}